\documentclass[12pt,a4paper,landscape]{iopart}
\usepackage[dvips]{color,graphics}
\begin{document}
\jl{1}

\title[Oscillator Neural Networks]{Analysis of Oscillator Neural
Networks for Sparsely Coded Phase Patterns}

\author{Masaki Nomura\footnote{E-mail:nomura@amp.i.kyoto-u.ac.jp} 
and Toshio Aoyagi}

\address{Department of Applied Mathematics and Physics, Graduate School
of Informatics, Kyoto University, Kyoto 606-8501, Japan\\
CREST, Japan Science and Technology, Kawaguchi, Saitama 332-0012, Japan
}

\begin{abstract}
We study a simple extended model of oscillator neural networks capable of
storing sparsely coded phase patterns, in which information is encoded both in 
the mean firing rate and in the timing of spikes. Applying the methods of
statistical neurodynamics to our model, we theoretically investigate the 
model's associative memory capability by evaluating its maximum storage
capacities and deriving its basins of attraction. It is shown that, as in
the Hopfield model, the storage capacity diverges as the activity level
decreases. We consider various practically and theoretically important cases.
For example, it is revealed that a dynamically adjusted threshold mechanism 
enhances the retrieval ability of the associative memory.
It is also found that, under suitable conditions, the network can recall
patterns even in the case that patterns with different activity levels are
stored at the same time. In addition, we examine the robustness with respect
to damage of the synaptic connections. The validity of these theoretical
results is confirmed by reasonable agreement with numerical simulations.
\end{abstract}



\section{Introduction}
One of the important but unsolved problems in neuroscience is to
determine how information is coded in neuronal activities. Most of the
traditional neural network models consisting of binary units are constructed
on the assumption that information is coded only in the mean firing rate of
the neurons. Although these models have provided us with theoretically
interesting information, they ignore many dynamical aspects of neuronal
activities. In fact, oscillatory activity appears to be ubiquitous in many
neuronal systems. For example, some recent biological experiments have revealed
that spatially synchronous oscillations of neuronal ensembles are dependent on
the global properties of the external stimulus. It has been suggested that
such synchronization is computationally significant in information processing
\cite{gray}. 
The hippocampus is also one of the areas in which neuronal
synchronization is observed and is believed to play an important role in memory
processing. 

With these new findings, many models concerning the dynamical aspects of neurons
and memory processing have been proposed \cite{malsburg,sompolinsky,terman}.
Among these, models consisting of networks of oscillator components are
particularly attractive, owing to their mathematical tractability. 
Like the Hopfield model, the simplicity of these models allows us to obtain
useful analytic results.
In fact, it has been reported that under suitable conditions, the
oscillator network models model associative memory, and we can
theoretically evaluate their maximum storage capacities and basins of
attraction \cite{abbot,cook,aoyagi1,kitano1,aoyagi2,aoyagi3}. 

However, such phase oscillator models are based on the unrealistic
situation that all the components in the network are always in the firing
state. In fact, real neurons can be in either the firing or non-firing state.
In addition, it is known that the level of activity in our brain is very
low. (i.e., at any given time only a small percentage of neurons are in the
firing state.) This situation is termed {\it sparse coding}.
From the results of theoretical studies of associative memory with binary units,
it has been found that the storage capacity diverges as $-1/a\ln a$ as the
firing rate $a$ becomes small
\cite{willshaw,tsodyks,vicente,buhmann,okada1,gardner}.

It thus seems that to faithfully to capture the essential dynamics of real
oscillatory neuronal systems, it is necessary to extend the oscillator model to
treat the non-firing state as well as the firing state. In this paper,
we present a simple extended oscillator neural network model of this nature.
Using this extended model, we study its associative memory ability to recall
sparsely coded phase patterns in which some neurons are in the non-firing
state and others encode information in the timings of spikes.

In the next section, we first review the theoretical basis of the phase
oscillator model and propose an extend version of the oscillator model to
treat non-firing states.
In the analysis of this model, we estimate the maximum storage capacity,
derive the basin of attraction, and evaluate the quality of recalled memories.
Then, we find that when we define the threshold as a dynamical variable in a
certain manner, the size of the basin of attraction can be increased.
Embedding patterns with different activity levels and the influence of 
synaptic dilution are also studied.

\section{Theories of Oscillator Neural Networks}
\label{onn}
\subsection{Phase Oscillator Model}
Let us first start with a survey of the theoretical basis of the phase
oscillator model. We assume that neurons fire periodically and interact
weakly with each other. In general, although such a neural system can be
described in terms of many internal dynamical variables, it is well known
that it can be reduced to a system of simple coupled phase oscillators
\cite{erementrout,kuramoto}. In this form, we can characterize the state
of the i-th neuron by a single variable, $\phi _i(t)$, which is referred
to as the {\it phase}. This variable represents the timing of the neuronal
spikes at time $t$. A typical reduced equation takes the form
\begin{equation}
  \frac{d\phi_i(t)}{dt}=\omega _i(t)+\sum_{j=1}^NJ_{ij}\sin(\phi_j(t)-\phi_i(t)
  +\beta_{ij}),
  \label{reduce}
\end{equation}
where $J_{ij}$ and $\beta_{ij}$ characterize the interaction between the i-th 
and j-th neurons. Assuming that all natural frequencies $\omega
_i(t)$ are equal to some fixed value $\omega _0$, we can eliminate the
$\omega_i(t)$ term in Eq.(\ref{reduce}) by redefining $\phi_i(t)$ through 
$\phi_i(t) \rightarrow \phi_i(t)+\omega_0t$. When we represent the state of
the i-th neuron by the complex form $W_i(t)=\exp(i\phi_i(t))$, Eq.(\ref{reduce})
can be written in the alternative form
\begin{equation}
  \frac{dW_i(t)}{dt}=\frac12(h_i(t)-\tilde{h_i}(t)W_i^2(t)),\qquad 
  h_i(t)=\sum_{j=1}^NC_{ij}W_j(t),
\end{equation}
where the complex variable $C_{ij}=J_{ij}\exp(i\beta_{ij})$ represents
the effect of the interaction between the i-th and j-th neurons and
$\tilde{h_i}(t)$ is the complex conjugate of $h_i(t)$. Considering the
fact that all neurons relax toward the equilibrium state satisfying the
relation $W_i=h_i/|h_i|$, we can simplify the above to 
the following discrete time system:
\begin{equation}
  W_i(t+1)=\frac{h_i(t)}{|h_i(t)|},\quad h_i(t)=\sum_{j=1}^NC_{ij}W_j(t).
  \label{original}
\end{equation}
This is known as a phase oscillator network model. It can be thought of as a
synchronous updated version of the phase oscillator neural network.

\subsection{Extended Model}
\label{model1}
The weakness of the model described by Eq.(\ref{original}) is that it can only
be used to treat the firing state. We wish to extend this model so that it has
the ability of retrieving sparsely coded phase patterns. In Eq.(\ref{original}),
$h_i(t)$ can be regarded as the local field produced by all other neurons.
This field determines the state of the i-th neuron
at the next time step. In a real neuron, when the membrane voltage is
less than some threshold, the neuron generates no neuronal spikes. 
Therefore, it is reasonable to extend the above model so that the generation
of spikes depends on the strength of the local field. 
Based on this consideration, for oscillator neural networks we propose
the following generalized model:
\begin{equation}
  W_i(t+1)=f(|h_i(t)|)\frac{h_i(t)}{|h_i(t)|}, \qquad
  h_i(t)=\sum_{j\ne i}^NC_{ij}W_j(t).
  \label{W_i}
\end{equation}
In this paper, we assume that $f(x)=\Theta(x - H)$, where $\Theta(x)$
is a step function defined by 
\begin{equation}
  \Theta(x)=
  \cases{
    1 & for $x \geq 0$ \cr
    0 & for $x < 0$, \cr
    }
\end{equation}
and $H$ is a threshold parameter controlling the activity of the network. 

\begin{figure}
  \begin{center}
    \scalebox{.5}{\includegraphics{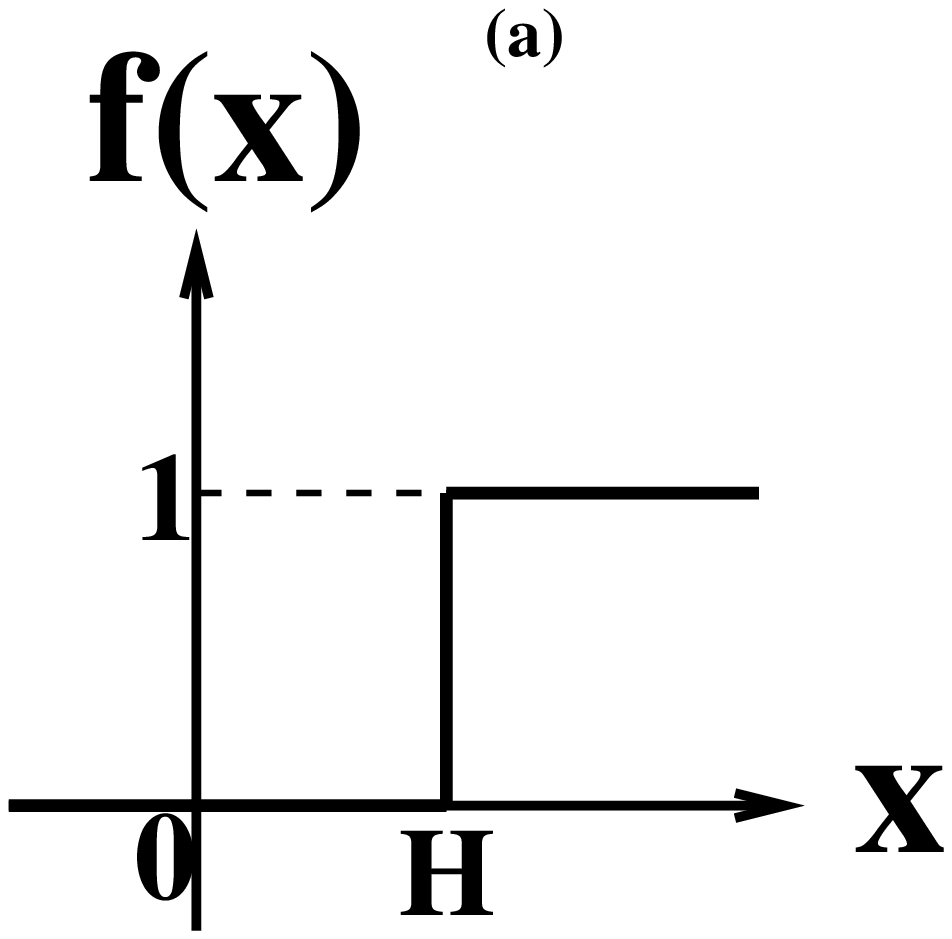}}
    \scalebox{.4}{\includegraphics{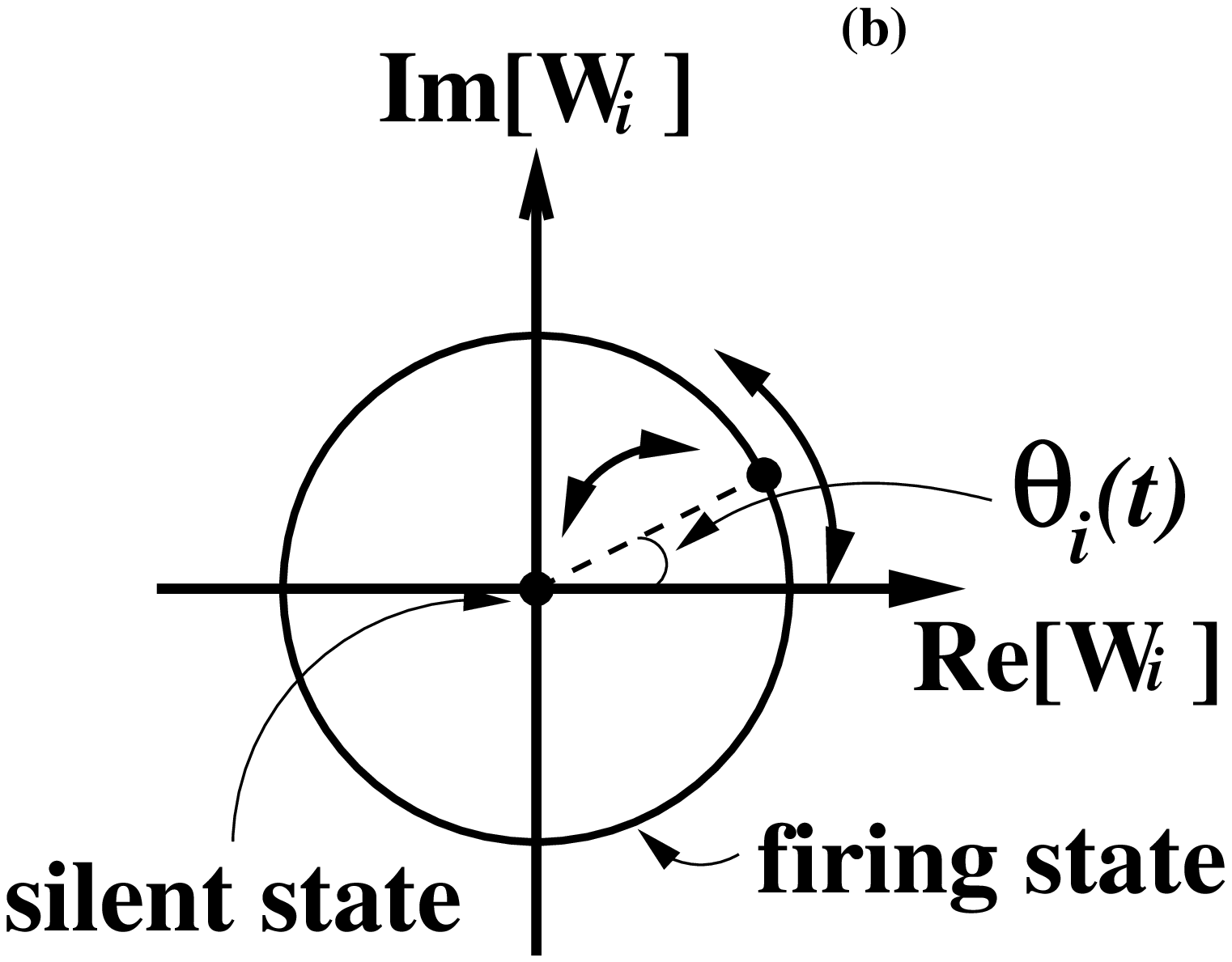}}
    \caption{(a) The step function $f(x)=\Theta(x - H)$, where $H$ is the 
      ``threshold'':
      When $|h_i(t)|\geq H$, $|W_i(t+1)|=1$, and otherwise, $|W_i(t+1)|=0$. 
      (b) Depiction of the dynamical change of the neuronal state. The circle
      and the origin 
      correspond to the firing state and the non-firing state, respectively.
      The phase and
      amplitude of $W_i(t)$ change according to Eq.(\ref{W_i}).\label{fw}} 
  \end{center}
\end{figure}

Figure\ref{fw}(a) displays the function $\Theta(x - H)$, and (b) illustrates
the dynamical change at updating. Since the amplitude $|W_i(t+1)|$ depends
on the threshold $H$, it has a strong influence on the activity of the system.

Now, we define a set of complex patterns denoted by $\xi_i^{\mu}=A_i^{\mu}
\exp(i\theta_i^{\mu})$ $(\mu =1,\cdots,P$ ; $i = 1,\cdots,N)$,
where $P$ is the total number of patterns and $N$ is the total number of
neurons. The variables $\theta_i^\mu$ and $A_i^{\mu}$ represent the phase 
and the amplitude of the i-th neuron 
in the $\mu$-th pattern, respectively. For theoretical simplicity, we choose
$A_i^{\mu}$ independently with the probability distribution:
\begin{equation}
  A_i^{\mu}=
  \cases{
    1 & for firing state with probability $a$\cr
    0 & for silent state with probability $1-a$.\cr
    }
\end{equation}
For the firing state,  $\theta_i^\mu$ is chosen at random from a uniform
distribution between $0$ and $2\pi$. Note that all the patterns 
have the same mean firing rate, $a$.
As the learning rule, we adopt a generalized Hebb rule taking the form
\begin{equation}
  C_{ij} = \frac1{aN}\sum_{\nu=1}^P\xi_i^{\nu}\tilde{\xi_j^{\nu}}.
\label{cij}
\end{equation}
This relation is based on information obtained from experiments on biological
systems that when the i-th and j-th neurons are firing simultaneously, the
connection between them is enhanced, and otherwise no modification occurs.

\subsection{Theoretical Analysis}
To analyze the recalling process theoretically, we must introduce several
macroscopic order parameters. The load parameter $\alpha$ is defined by
$\alpha = P/N$. The overlap $M^\mu(t)$ between $W_i(t)$ and $\xi_i^\mu$ at
time $t$ is defined by
\begin{equation}
  M^{\mu}(t)=m^{\mu}(t)\exp(i\varphi^{\mu}(t))=\frac1{aN}\sum_{j=1}^N
  \tilde{\xi}_j^{\mu}W_j(t).
  \label{M_i}
\end{equation}
In practice, owing to rotational symmetry, the similarity between the state
of the system $W_i(t)$ and the $\mu$-th pattern $\xi_i^{\mu}$ can be measured 
by $m^{\mu}(t)$, the amplitude of $M^{\mu}(t)$.
Now, let us consider the situation in which the system is retrieving the 
pattern $\xi_i^1$, that is,
\begin{equation}
  m(t)\equiv m^1(t)\sim O(1),\qquad m^{\mu}(t)\sim O\left(\frac1{\sqrt{N}}\right)
  \quad(\mu\ne 1).\label{order}
\end{equation}
The local field $h_i(t)$ can be separated as
\begin{equation}
  h_i(t)=M^1(t)\xi_i^1+z_i(t)=|\xi_i^1|m(t)e^{i(\varphi^1(t)+\theta_i^1)}+
  z_i(t),
  \label{h_i}
\end{equation}
where $z_i(t)$ is defined by
\begin{equation}
  z_i(t)=\frac1{aN}\sum_{j,\nu\ne1}\xi_i^{\nu}\tilde{\xi_j^{\nu}}W_j(t).
  \label{z_i}
\end{equation}
The first term in Eq.(\ref{h_i}) is the signal driving the system to
recall the pattern, and the second term can be regarded as the noise
arising from the other learned patterns. It is the essence of our
analysis that we treat the second noise term as a complex Gaussian noise
characterized by
\begin{equation}
  <z_i(t)>_i=0,\quad<|z_i(t)|^2>_i=2\sigma^2(t).
  \label{sigma}
\end{equation}
Under the above assumptions, we study the properties of this
network by applying the methods of statistical
neurodynamics \cite{amari2,fukai,okada2,coolen}. 
To begin with, we calculate the overlap at time $t+1$.
From Eqs.(\ref{M_i}), (\ref{W_i}) and (\ref{h_i}), $M^1(t+1)$ is given by
\begin{eqnarray}
  M^1(t+1)&=&m(t+1)\exp(i \varphi ^1(t+1))=\frac1{aN}\sum_{j=1}^N
  \tilde{\xi}_j^1f(|h_j(t)|)\frac{h_j(t)}{|h_j(t)|}\nonumber\\
  &=&\frac1{aN}\sum_j\tilde{\xi_j^1}f(|\xi_j^1M^1(t)+z_j(t)|)
  \frac{\xi_j^1M^1(t)+z_j(t)}{|\xi_j^1M^1(t)+z_j(t)|}\nonumber.
\end{eqnarray}
Now, we assume that the phase of $M^1(t+1)$ is almost constant, that is,
$\varphi^1(t+1)\approx\varphi_0$. 
The validity of this assumption is supported by the results of preliminary
numerical simulations. Owing to the rotational symmetry of the complex
Gaussian noise, we can replace $z_j(t)$ with
$z_j(t)\exp[i(\varphi_0+\theta^1_j)]$. After some calculations,
in the limit $N \rightarrow \infty$, we obtain
\begin{eqnarray}
  m(t+1)&=&\frac1{aN}\sum_j|\xi_j^1|f(||\xi_j^1|m(t)+z_j(t)|)
  \frac{|\xi_j^1|m(t)+z_j(t)}{||\xi_j^1|m(t)+z_j(t)|}
  \nonumber\\
  &=&\left<\left<f(|m(t)+z(t)|)\frac{m(t)+z(t)}
      {|m(t)+z(t)|}\right>\right>_{z(t)},
  \label{m}
\end{eqnarray}
where $\ll\qquad\gg_{z(t)}$ represents the average over the complex Gaussian
noise $z(t)$ defined by Eq.(\ref{sigma}). To calculate Eq.(\ref{m}) numerically, 
we need the value of $\sigma(t)$, which is the variance of $z(t)$.
Thus, in the next step, we consider the relation between
$z_i(t+1)$ and $z_i(t)$, from which we can obtain the relation between
$\sigma(t+1)$ and $\sigma(t)$.
From Eq.(\ref{z_i}), the noise at time $t+1$ can be written as
\begin{eqnarray}
  z_i(t+1)&=&\frac1{aN}\sum_{j,\nu\ne1}\xi_i^{\nu}\tilde{\xi_j^{\nu}}
  W_j(t+1)\nonumber\\
  &=&\frac1{aN}\sum_{j,\nu\ne1}\xi_i^{\nu}\tilde{\xi_j^{\nu}}
  f(|h_j(t)|)\frac{h_j(t)}{|h_j(t)|}\nonumber\\
  &=&\frac1{aN}\sum_{j,\nu\ne1}\xi_i^{\nu}\tilde{\xi_j^{\nu}}
  f(|h^{-\nu}_j(t)+h^{\nu}_j(t)|)\frac{h^{-\nu}_j(t)+
    h^{\nu}_j(t)}{|h^{-\nu}_j(t)+h^{\nu}_j(t)|},
  \label{z}
\end{eqnarray}
where $h^{-\nu}_j(t)$ and $h^{\nu}_j(t)$ are defined by
\begin{equation}
  h^{-\nu}_j(t)=\frac1{aN}\sum_{k,\mu\neq\nu}\xi_j^{\mu}\tilde{\xi_k^{\mu}}
  W_k(t),\quad
  h^{\nu}_j(t)=\frac1{aN}\sum_k\xi_j^{\nu}\tilde{\xi_k^{\nu}}W_k(t).
\end{equation}
Note that these functions satisfy the relation
$h_j(t)=h^{-\nu}_j(t)+h^{\nu}_j(t)$. We can carry out the summation in
Eq.(\ref{z}) under the assumption that $h^{-\nu}_j(t)$ is independent
of $\xi_i^{\nu}$. Doing so, we obtain
\begin{eqnarray}
  z_i(t+1)&\simeq&K(m(t),\sigma(t))+z_i(t)G(m(t),\sigma(t))
  \label{z2}\\
  K(m(t),\sigma(t))&=&\frac1{aN}\sum_{j,\nu\ne1}\xi_i^{\nu}
  \tilde{\xi_j^{\nu}}f(|h_i(t)|)\frac{h_i(t)}{|h_i(t)|}
  \label{k}\\
  G(m(t),\sigma(t))&=&\left<\left<a\left(\frac{f^{\prime}(|m(t)+z(t)|)}2+
        \frac{f(|m(t)+z(t)|)}{2|m(t)+z(t)|}\right)\right.\right.
  \label{g}\\
  &&\left.\left.+(1-a)\left(\frac{f^{\prime}(|z(t)|)}2+\frac{f(|z(t)|)}{2|z(t)|}
      \right)\right>\right>_{z(t)}.\nonumber
\end{eqnarray}
Note that Eqs.(\ref{z2}), (\ref{k}) and (\ref{g}) are needed to calculate
$\sigma$ in both the equilibrium and the non-equilibrium cases we will 
describe in the next two sections.

\subsection{Equilibrium State}
In this section, we consider the equilibrium state of our model,
in which $z_i(t)$ is constant. Applying $z_i(t+1)
=z_i(t)=z_i$ to Eq.(\ref{z2}), we obtain
\begin{equation}
  z_i=\frac{K}{1-G}.
\end{equation}
Using this equation, we immediately obtain
\begin{equation}
  2\sigma^2=<|z_i|^2>_i=\frac{|K|^2}{(1-G)^2},
\end{equation}
where
\begin{eqnarray} 
  |K|^2&=&\alpha\left<\left<af(|m+z|)
      +(1-a)f(|z|)\right>\right>_{z}\\
  &\equiv&\alpha Q.
\end{eqnarray}
Consequently, we find that the equilibrium state satisfies the equations
\begin{eqnarray}
  m&=&\left<\left<f(|m+z|)\frac{m+z}{|m+z|}\right>\right>_{z}
  \label{mconst}\\
  \sigma^2&=&\frac{\alpha}{2(1-G)^2}Q\label{sconst}\\
  G&=&\left<\left<a\left(\frac{f^{\prime}(|m+z|)}2
        +\frac{f(|m+z|)}{2|m+z|}\right)\right.\right.
  \label{gconst}\\
  &&\left.\left.+(1-a)\left(\frac{f^{\prime}(|z|)}2+\frac{f(|z|)}{2|z|}
      \right)\right>\right>_z
  \nonumber\\
  Q&=&\left<\left<af(|m+z|)+(1-a)f(|z|)\right>\right>_{z}.
  \label{qconst}
\end{eqnarray}

Solving the above equilibrium equations, we can find two types of solutions.
If the load parameter $\alpha$ is smaller than a certain value
$\alpha_c$, there exists a solution for which the overlap $m$ between the
system and the pattern is not zero. As $m\neq0$ implies that the system 
has a correlation with the retrieved pattern, this solution corresponds 
to the retrieval state. Then, if $\alpha$ is larger 
than $\alpha_c$, there is only one solution, and for this solution $m=0$.
This solution therefore corresponds to the non-retrieval state.

\begin{figure}
  \begin{center}
    \scalebox{.4}{\includegraphics{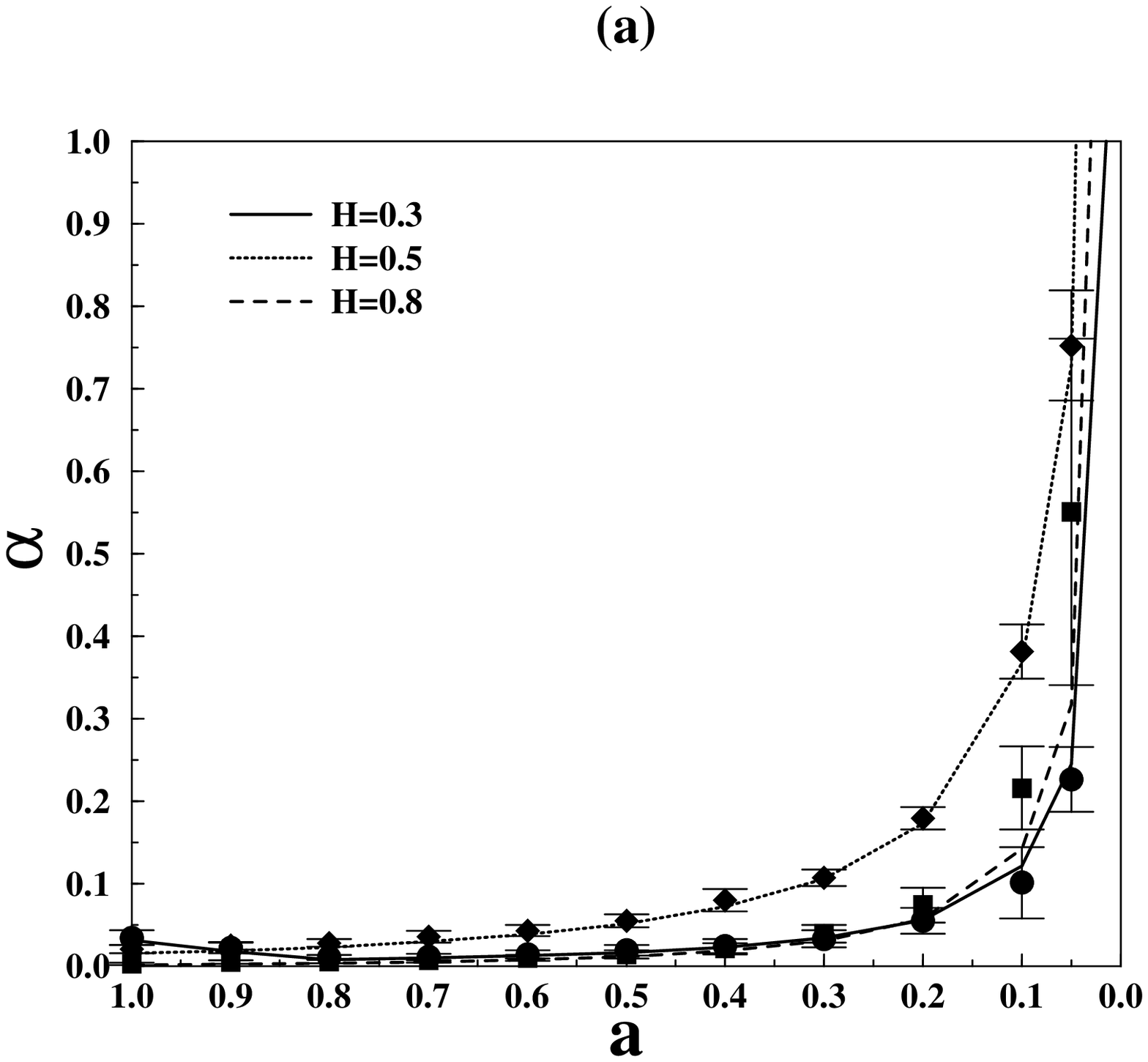}}
    \scalebox{.88}{\includegraphics{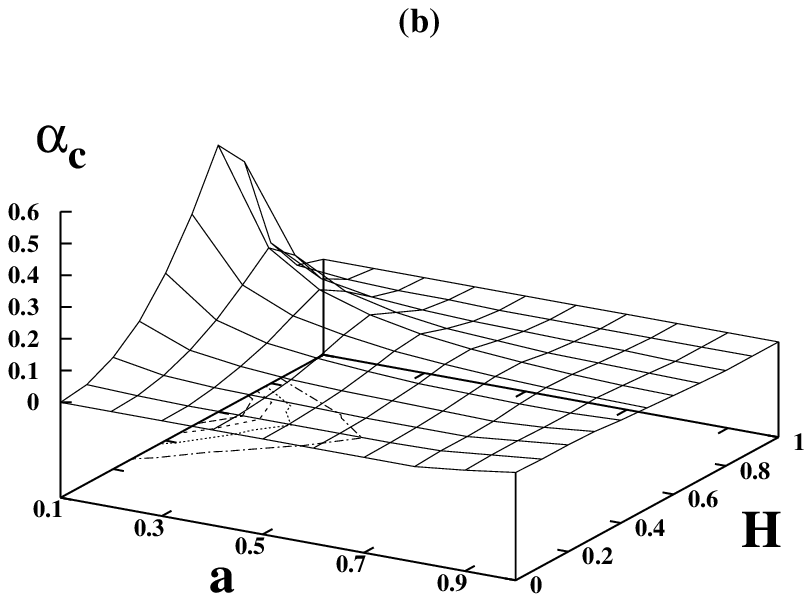}}
    \caption{(a) Storage capacity $\alpha_c$ as a function of the mean
      activity level $a$ for various thresholds $H$. The data points
      indicate numerical results with $N=1000$ for $20$ trials.
      The lines were obtained from theoretical analysis.
      (b) A three-dimensional isometric plot of the maximum storage capacity 
      $\alpha_c$ as a function of $a$ and $H$ from the theoretical results.
      \label{capacity2d}}
  \end{center}
\end{figure}

The critical load parameter
$\alpha_c$ is called ``the maximum storage capacity''.
We now examine our theoretical results by comparing them with results from
numerical simulations. In Fig.\ref{capacity2d}(a), we show the dependence
of $\alpha_c$ on several parameters, such as the threshold $H=0.3,0.5,0.8$
and the activity level $a$.
In (a), We can see that as the activity level $a$ decreases, the storage
capacity $\alpha_c$ increases for each $H$. In particular, in the limit
$a \rightarrow 0$, we numerically find that the storage capacity diverges
in proportion to $-1/a\ln a$, as in the case of the Hopfield model. 
The maximum storage capacity as a function of $a$ and $H$ is illustrated in (b).
It is shown there that for any $H$, the storage capacity diverges as $a$
decreases to zero.

\subsection{Non-equilibrium State}

In this section, we derive dynamical equations to study the retrieving process.
To obtain a recursion equation for $\sigma$, we start with
Eq.(\ref{z2}). Squaring Eq.(\ref{z2}), we obtain
\begin{eqnarray}
  2\sigma^2(t+1)&=&\alpha Q(m(t),\sigma(t))+2\sigma^2(t)G^2(m(t),\sigma(t))
  \nonumber\\
  &&+2Re\left<\left<K(m(t),\sigma(t))\tilde{z}_i(t)\tilde{G}(m(t),\sigma(t))
    \right>\right>_{z(t)},
\end{eqnarray}
where $K(m(t),\sigma(t))$ and $G(m(t),\sigma(t))$ are given by Eqs.(\ref{k})
and (\ref{g}), and
\begin{equation}
  Q(m(t),\sigma(t))=\left<\left<\right.\right.af(|m(t)+z(t)|)
  +(1-a)f(|z(t)|)\left.\left.\right>\right>_{z(t)}\label{q(t)}.
\end{equation}

To proceed with the calculation, we must estimate the time correlation of the
noise. For the first-order approximation, we ignore the temporal correlation of
$z_i(t)$. We then obtain
\begin{eqnarray}
  \sigma^2(t+1)&=&\frac{\alpha}2Q(m(t),\sigma(t))+\sigma^2(t)G^2(m(t),\sigma(t))
  \nonumber\\
  &&+\alpha a^2m(t+1)m(t)G(m(t),\sigma(t)).
  \label{1stsigma}
\end{eqnarray}
This corresponds to the Amari-Maginu theory in the case of traditional
neural networks.

For the second-order approximation, we take into account the fact that
$z_i(t)$ is correlated only with $z_i(t-1)$, while all correlations 
with $z_i(t^{\prime})$ for $t^{\prime}<t-1$ are ignored.
In this case, $\sigma(t+1)$ can be obtained from
\begin{eqnarray}
  \sigma^2(t+1)&=&\frac{\alpha}2 Q(m(t),\sigma(t))
  +\sigma^2(t)G^2(m(t),\sigma(t))\nonumber\\
  &&+\alpha G(m(t),\sigma(t))X(t+1,t)\\
  &&+\alpha a^2m(t+1)m(t-1)G(m(t),\sigma(t))G(m(t-1),\sigma(t-1)),\nonumber
\end{eqnarray}
where $Q(m(t),\sigma(t))$ is given by Eq.(\ref{q(t)}), $G(m(t),\sigma(t))$ is
given by Eq.(\ref{g}), and
\begin{equation}
  X(t+1,t)=Re<W_j(t+1)\tilde{W}_j(t)>_j\label{xt}.
\end{equation}
Here,
\begin{equation}
  \rho(t,t-1)=\frac{\alpha X(t,t-1)}{2\sigma(t)\sigma(t-1)}
  +\frac{\sigma(t-1)}{\sigma(t)}G(m(t-1),\sigma(t-1))\label{ro}
\end{equation}
is needed to evaluate Eq.(\ref{xt}). Note that $\rho(t,t-1)$ is the correlation 
coefficient between $z(t)$ and $z(t-1)$.
For the second order approximation, consequently, the retrieving process
of the network is described by the equations
\begin{eqnarray}
  m(t+1)&=&\frac1{aN}\sum_j|\xi_j^1|f(||\xi_j^1|m(t)+z_j(t)|)
  \frac{|\xi_j^1|m(t)+z_j(t)}{||\xi_j^1|m(t)+z_j(t)|}\nonumber\\
  &=&\left<\left<f(|m(t)+z(t)|)\frac{m(t)+z(t)}
      {|m(t)+z(t)|}\right>\right>_{z(t)}\\
  \sigma^2(t+1)&=&\frac{\alpha}2 Q(m(t),\sigma(t))
  +\sigma^2(t)G^2(m(t),\sigma(t))\nonumber\\
  &&+\alpha G(m(t),\sigma(t))X(t+1,t)\label{2ndsigma}\\
  &&+\alpha a^2m(t+1)m(t-1)G(m(t),\sigma(t))G(m(t-1),\sigma(t-1))\nonumber\\
  X(t+1,t)&=&Re<W_j(t+1)\tilde{W}_j(t)>_j\\
  \rho(t,t-1)&=&\frac{\alpha X(t,t-1)}{2\sigma(t)\sigma(t-1)}
  +\frac{\sigma(t-1)}{\sigma(t)}G(m(t-1),\sigma(t-1))\\
  G(m(t),\sigma(t))&=&\left<\left<a\left(\frac{f^{\prime}(|m(t)+z(t)|)}2
        +\frac{f(|m(t)+z(t)|)}{2|m(t)+z(t)|}\right)\right.\right.\label{2ndg}\\
  &&\left.\left.+(1-a)\left(\frac{f^{\prime}(|z(t)|)}2+\frac{f(|z(t)|)}{2|z(t)|}
      \right)\right>\right>_{z(t)}\nonumber\\
  Q(m(t),\sigma(t))&=&\left<\left<\right.\right.af(|m(t)+z(t)|)
  +(1-a)f(|z(t)|)\left.\left.\right>\right>_{z(t)}\label{2ndq}.
\end{eqnarray}

For initial conditions, we choose $\sigma^2(0)=a\alpha/2$, $X(0,-1)=0$ 
and $X(1,0)=a^2 m(1)m(0)$. We choose many values of the initial overlap $m(0)$
and carry out numerical calculations for each. In this way, we determine the
lower bound, under which the network fails to retrieve the pattern.

\begin{figure}
  \begin{center}
    \scalebox{.4}{\includegraphics{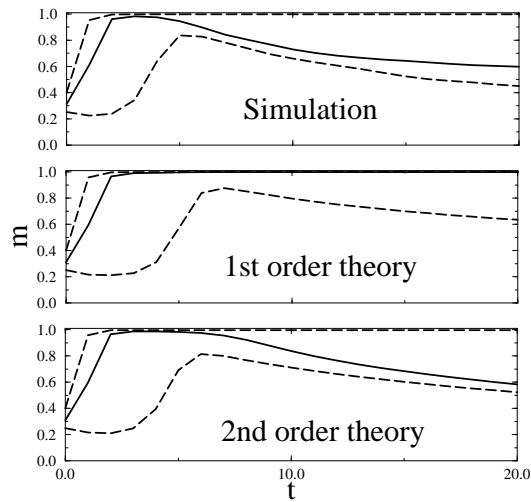}}
    \caption{Time evolution of the overlap $m(t)$ for various initial
      conditions: $m=0.25,0.31,0.4$. Here $\alpha=0.013$, $H=0.3$ and
      $a=0.5$. Results of the second-order approximation give better agreement
      with the numerical simulation than those of the first-order
      approximation.\label{timestep}}
  \end{center}
\end{figure}

For higher-order approximations, we could derive similar
generalized equations describing the retrieval dynamical process. However,
as shown in Fig.\ref{timestep}, the theoretical prediction at second
order gives reasonable agreement with the numerical simulation in contrast to
that at first order. We thus conclude that it is sufficient to use
the second-order approximation for the theoretical analysis in this paper.

\begin{figure}
  \begin{center}
    \scalebox{.4}{\includegraphics{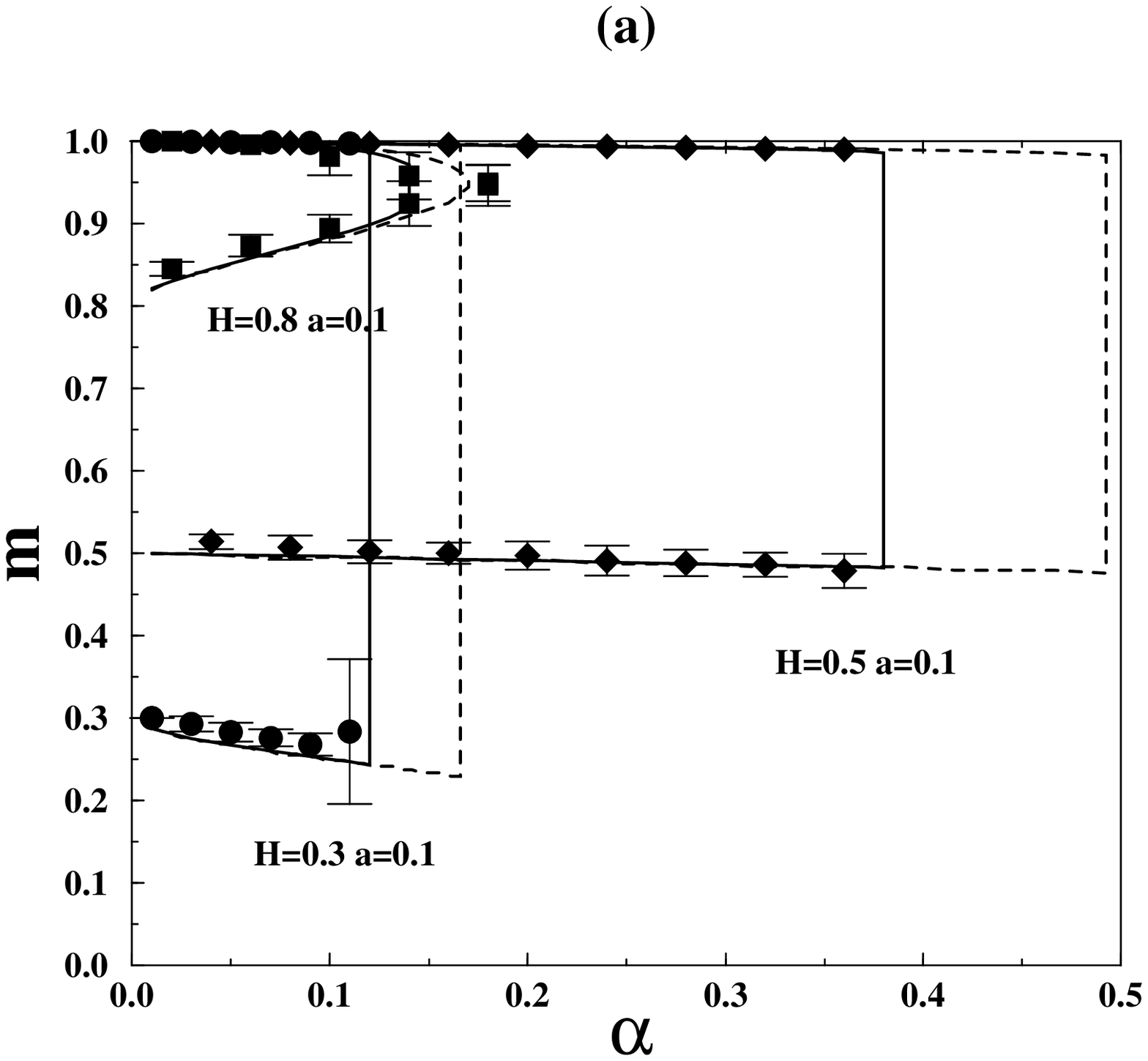}}
    \scalebox{.4}{\includegraphics{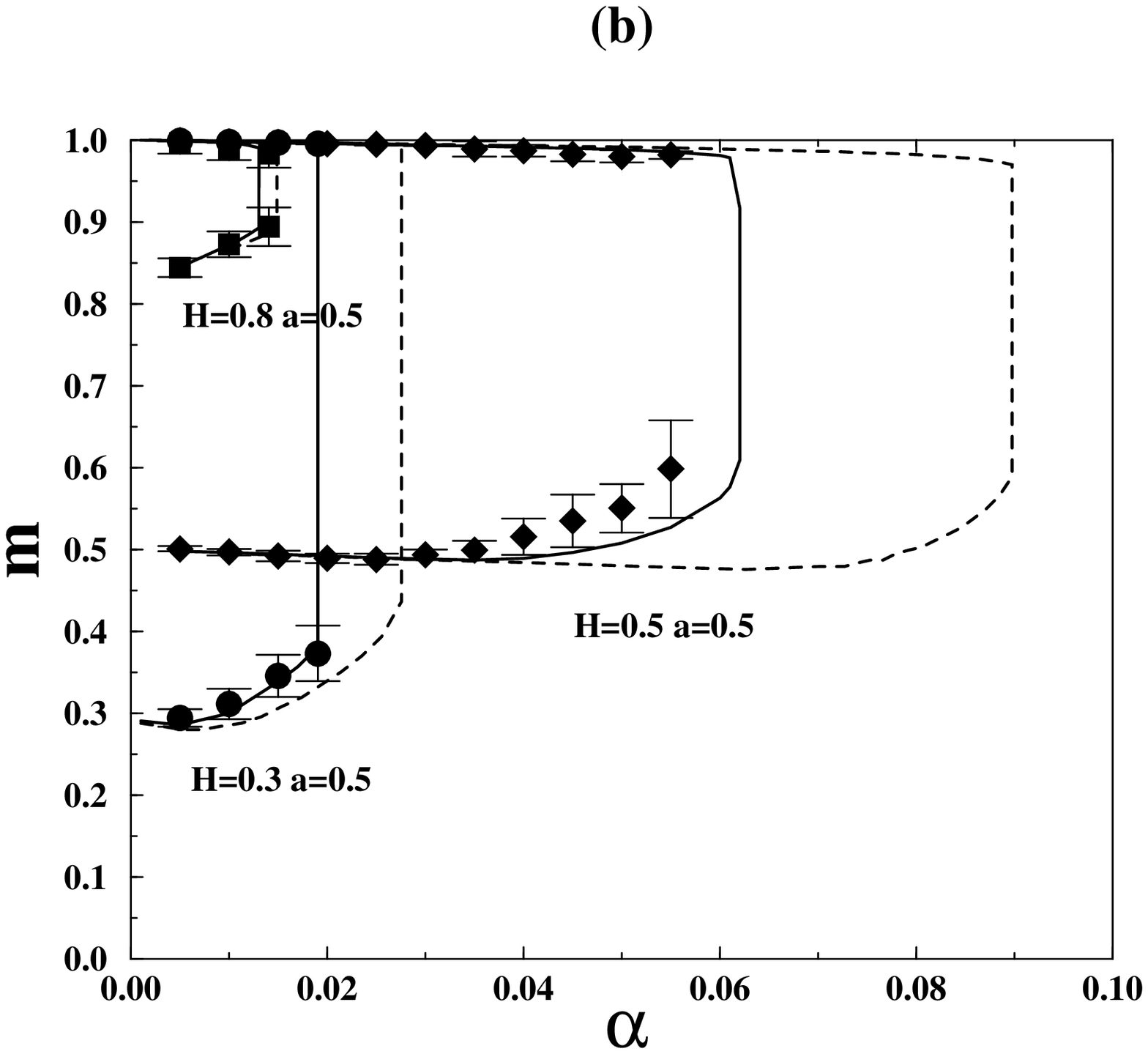}}
    \caption{(a) Basins of attraction for various values of the threshold $H$, 
      with $a=0.1$.
      The data points are the results of the numerical simulation, the dashed 
      lines correspond to the first-order approximation, and the solid lines 
      correspond to the 
      second-order approximation. The results of the numerical simulation were 
      obtained with
      $N=1000$ for $20$ trials.
      (b) Same as (a) but with $a=0.5$.\label{basin1}}
  \end{center}
\end{figure}

Figure\ref{basin1} displays a phase diagram for $m(0)_c$ and $m(\infty)$
at various mean activity levels $a$ and thresholds $H$. The
solid lines and the data points here correspond to the theoretical and the
numerical results, respectively. It is seen that even in the region satisfying
$\alpha < \alpha_c$, the system cannot retrieve the pattern if the initial
overlap $m(0)$ is smaller than $m(0)_c$.
The boundary corresponding to $m(0)_c$ is represented by the lower curves in
Fig.\ref{basin1}. Thus, for $m(0)\geq m(0)_c$, $m(t)$ reaches the value of
the upper curves $m(\infty)$, while for $m(0)<m(0)_c$, $m(t)$ decreases to zero. 
We should note that the basins of attraction remain wide even near $\alpha_c$.
This can be thought of as an advantage of associative memory. 

To predict the correct behavior of the retrieval dynamics, it is necessary
for the time correlation of the noise terms to be taken into account,
and thus the second-order approximation discussed above is necessary.
In the case we consider, this order is also sufficient, but interestingly,
it has been reported that the fourth-order approximation is necessary
when $a=1$ and $H=0$ \cite{aoyagi1}. 
The question arises why the second-order approximation is not sufficient
in this case, while it is sufficient in the case we consider.
The key point here is that the order of the last term in Eq.(\ref{2ndsigma}) is 
proportional to the square of the activity level $a^2$.
For $a<1$, this factor $a^2$ weakens the influence of the time correlation 
on the recalling process. Consequently, for a sparse coded pattern with 
$a<1$, even the second-order approximation results in reasonable agreement 
with the numerical results.

\subsection{Sequence Generator}
In this section, we consider the case in which the network retrieves a cyclic
sequence of $P$ patterns associatively, say $\xi^1 \rightarrow \xi^2 \rightarrow
\cdots \rightarrow \xi^P \rightarrow \xi^1 \rightarrow \cdots$.
In order to allow for such a process, we employ synaptic connections of the form
\begin{equation}
  C_{ij}=\frac1{aN}\sum_{\nu=1}^P\xi_i^{\nu+1}\tilde{\xi_j^{\nu}}.
\end{equation}
In a manner similar to that in the derivation of Eqs.(\ref{1stsigma}) and
(\ref{2ndsigma}), we obtain the following equations: 
\begin{eqnarray}
  \sigma^2(t+1)&=&\frac{\alpha}2 Q(m(t),\sigma(t))+\sigma^2(t)G^2(m(t),\sigma(t))
  \label{secsigma}\\
  m(t+1)&=&\left<\left<f(|m(t)+z(t)|)\frac{m(t)+z(t)}{|m(t)+z(t)|}
    \right>\right>_{z(t)}.
\end{eqnarray}
Note that, since the target pattern changes from time 
to time, for the definition of the overlap we adopt $m(t)\equiv m^\mu(t)=
|\frac1{aN}\sum_{j=1}^N\tilde{\xi_j^\mu}W_j(t)|$, where $\mu$ is the number
of the target pattern at time $t$. Here $G(m(t),\sigma(t))$ and
$Q(m(t),\sigma(t))$ are the same as those defined by Eqs.(\ref{2ndg})
and (\ref{2ndq}), respectively. We should note that in the limit
$N\rightarrow\infty$, the last term in Eqs.(\ref{1stsigma}) and (\ref{2ndsigma})
vanishes in Eq.(\ref{secsigma}), because the effect of the time correlation can
be ignored in the above derivation. Therefore, in a sequence generator,
it is expected that our theoretical prediction is almost exact.

\begin{figure}
  \begin{center}
    \scalebox{.4}{\includegraphics{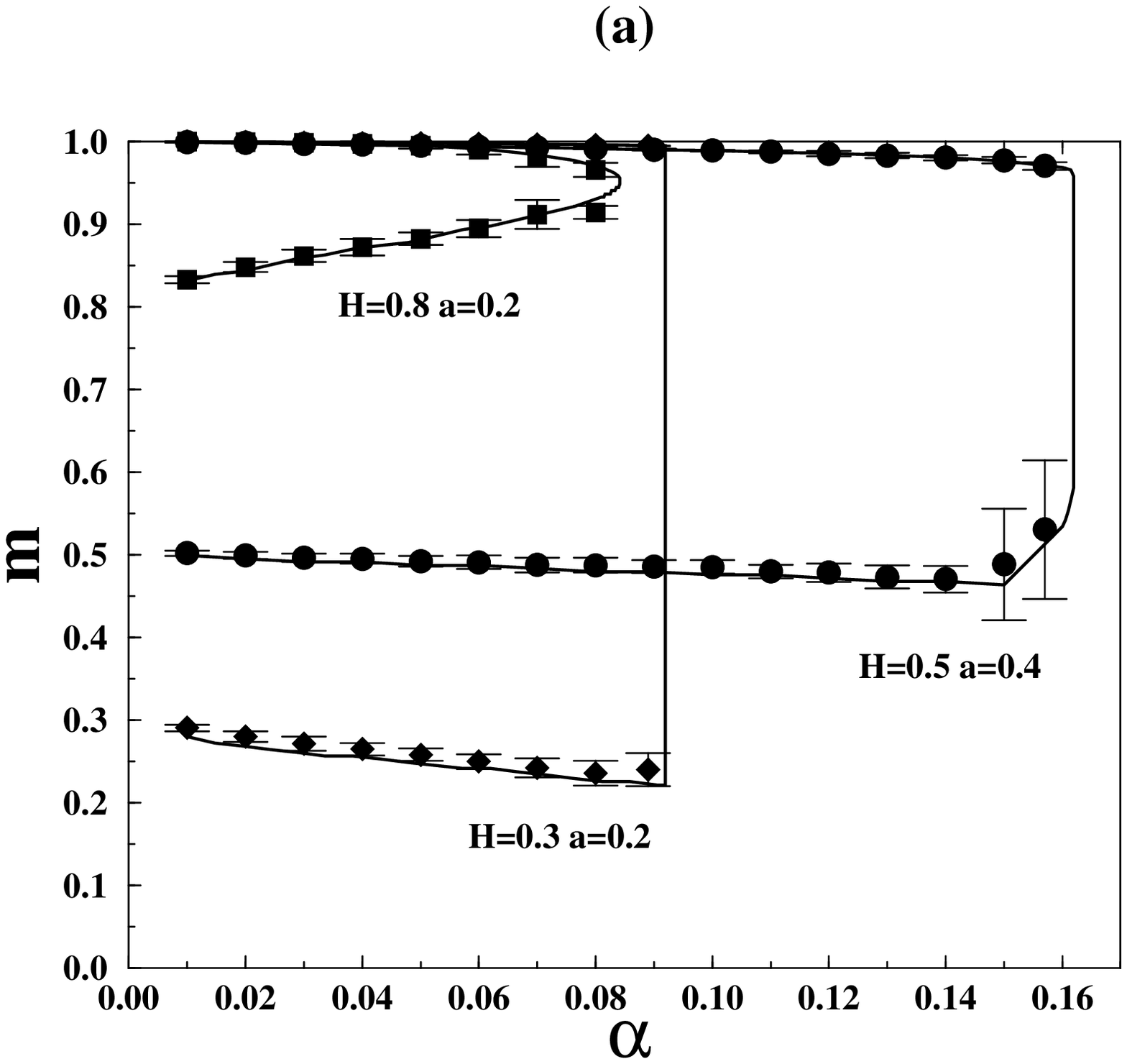}}
    \scalebox{0.88}{\includegraphics{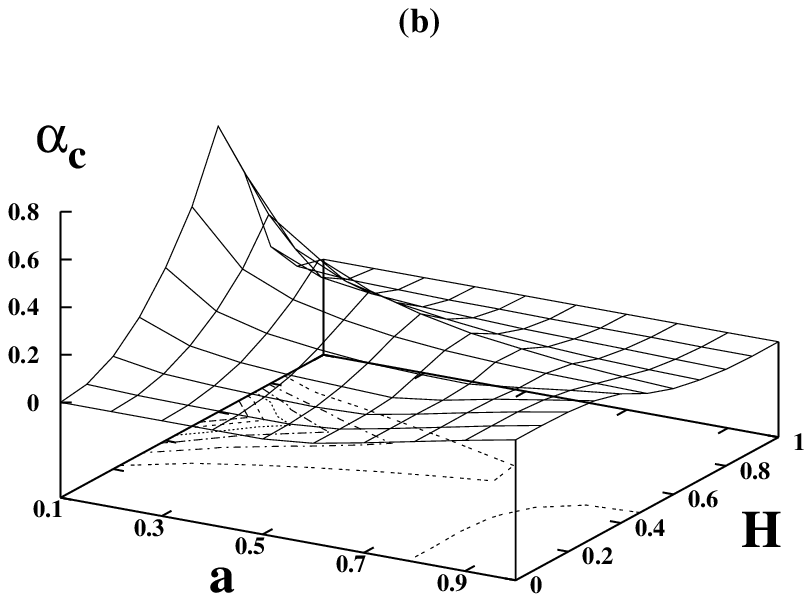}}
    \caption{(a) Basins of attraction for various firing rates $a$ and 
      thresholds
      $H$. The data points are the results of the numerical simulations 
      ($N=1000$ for $20$ trials), and the solid lines are the theoretical 
      results. (b) A three-dimensional isometric plot of the
      maximum storage capacity $\alpha_c$ as a function of $a$ and $H$ from 
      the theoretical results.
      \label{basinseq}}
  \end{center}
\end{figure}

Figure\ref{basinseq}(a) displays a phase diagram obtained from our theoretical
analysis. All the lines of theoretical results agree with the numerical ones 
quite well, as expected. This agreement suggests that higher-order
approximations will result in even better agreement and leads us to believe
that our theoretical derivation is valid. In (b), it is also shown that,
like auto-associative memory, the storage capacity diverges in the limit
$a \rightarrow 0$. This is regarded as expressing the meaning that in the
limit of sparse coding (i.e., as $a \rightarrow 0$),
we can embed an infinite length of sequential patterns.

\subsection{Dynamically Adjusted Threshold}

\begin{figure}
  \begin{center}
    \scalebox{.4}{\includegraphics{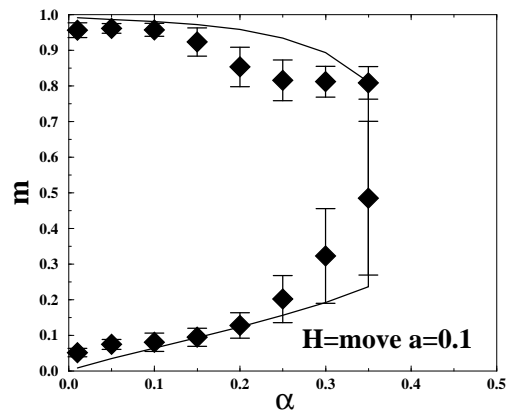}}
    \caption{Basins of attraction in the case that $H(t)$ is a dynamically 
      adjusted variable with $a=0.1$.
      Comparing with Fig.\ref{basin1}(a), we find that this mechanism enlarges
      the basin of attraction. The data points represent the results of a 
      numerical simulation with $N=1000$ for
      $20$ trials, and the line is the theoretical result.\label{basinself}}
  \end{center}
\end{figure}

From Fig.\ref{basin1}, it appears that the basin of attraction for our model is
smaller than those in the binary model and the phase oscillator. 
In this section, we attempt to increase the size of the basin of attraction
by defining the threshold as a dynamical variable that is proportional to
the standard deviation of the noise. Therefore, assuming the same condition
as in Sec.\ref{model1}, we add the equation
\begin{equation}
  H(t) = \sqrt{-2\ln a}\sigma(t),
\end{equation}
where $H(t)$ is the threshold at time $t$. This choice for the form of $H(t)$
is made to insure the relations $m\simeq 1$ and [firing rate] $\simeq a$,
which cause the Hamming distance between the state of the neuron and the
retrieved pattern to be small \cite{control}. Comparing Fig.\ref{basinself}
with Fig.\ref{basin1}, it is seen that the basin of attraction can be
enlarged by introducing the dynamically adjusted threshold into our model.
Therefore, when we introduce the dynamically adjusted threshold, our model
has an advantage similar to that of the binary model and the phase oscillator.

\section{Patterns Generated with Different Firing Rates}

To this point, we have assumed that the activity levels are equal for all
patterns. However, it is likely that the actual activity
level depends on the pattern which the network is retrieving presently,
in other words, on the content of the required information.
Unfortunately, using a traditional neural network, we encounter
difficulties in storing patterns with different activity levels simultaneously.
In this section, we demonstrate that, unlike traditional models, the proposed
oscillator model can easily store multiple patterns with
different activity levels using a simple Hebbian learning rule.

Let us consider a set of complex patterns defined by
\begin{equation}
  Prob[|\xi_i^\mu|=1]=
  \cases{
    a_1 & for $1\leq\mu\leq P_1$ \cr
    a_2 & for $P_1+1\leq\mu\leq P$, \cr
    }
\end{equation}
where generally $a_1\not=a_2$. 
Thus, the total number of the patterns with activity level $a_1$ is
$P_1$, while the total number of patterns with the activity level $a_2$ is
$P_2=P-P_1$. Using the above patterns, grouped into two different activity
levels, we examine whether the network can retrieve patterns having
different firing rates. To retrieve such patterns, we use the following
modified form of the learning rule Eq.(\ref{cij}):
\begin{equation}
  C_{ij} = \frac1{a_1N}\sum_{\nu=1}^{P_1}\xi_i^{\nu}\tilde{\xi_j^{\nu}}
  +\frac1{a_2N}\sum_{\nu=P_1+1}^{P}\xi_i^{\nu}\tilde{\xi_j^{\nu}}.
\end{equation}
Note that this is slightly different from the covariance rule adopted in the
context of the learning of sparsely coded patterns. Owing to the rotational
symmetry of the phase distribution in the patterns, $C_{ij}$ can be defined
in terms of the patterns $\xi_i^\nu$ themselves, rather than the difference
between the patterns and the average activity.

\begin{figure}
  \begin{center}
    \scalebox{.4}{\includegraphics{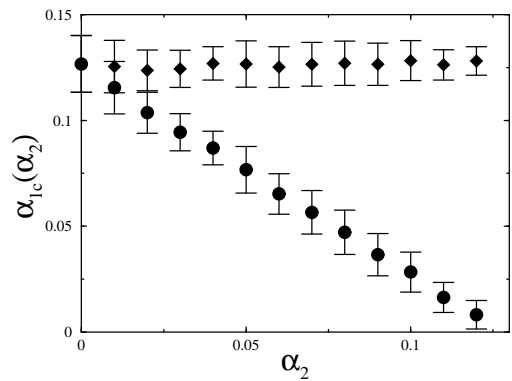}}
    \caption{Storage capacity $\alpha_{1c}$ as a function of $\alpha_2$ for
      $a_1=0.1$ and $a_2=0.2$. The circular data points represent $\alpha_{1c}$
      as a function of $\alpha_2$, and the diamond-shaped points represent
      $\alpha_{1c}+\alpha_2$. Both of them were obtained from a numerical
      simulation with $N=2000$ for $20$ trials. \label{doublecapa}}
  \end{center}
\end{figure}

With the method described in Sec.\ref{onn}, we can derive both equilibrium 
and dynamical equations. The results have the same forms as those in
Sec.\ref{onn}. However, here the value $a_1$ should be used in place of
the activity parameter $a$ in the previous equations, because the retrieval
pattern $\xi_i^1$ has an activity level $a_1$. Let us define the load parameter
as $\alpha_1=P_1/N$ with respect to the retrieval pattern.
(The usual load parameter is defined as $\alpha=P/N=(P_1+P_2)/N$.)
The theoretical analysis yields $\alpha_{1c}=const.-\alpha_2$, where
$\alpha_{1c}$ is the maximum storage capacity for $a_1$, and $\alpha_2$ is
the storage capacity for $a_2$. Figure\ref{doublecapa} displays $\alpha_{1c}$
as a function of $\alpha_2=P_2/N$. We can see from Fig.\ref{doublecapa} that
$\alpha_{1c}=\alpha(a_1)-\alpha_2$, where the constant value $\alpha(a_1)$
is given by the equations in (\ref{mconst}), (\ref{sconst}), 
(\ref{gconst}) and (\ref{qconst}) for $a=a_1$. 
For the maximum storage capacity $\alpha_{2c}$ associated with the 
activity level $a_2$, we have $\alpha_{2c}=\alpha(a_2)-\alpha_1$.

\begin{figure}
  \begin{center}
    \scalebox{.4}{\includegraphics{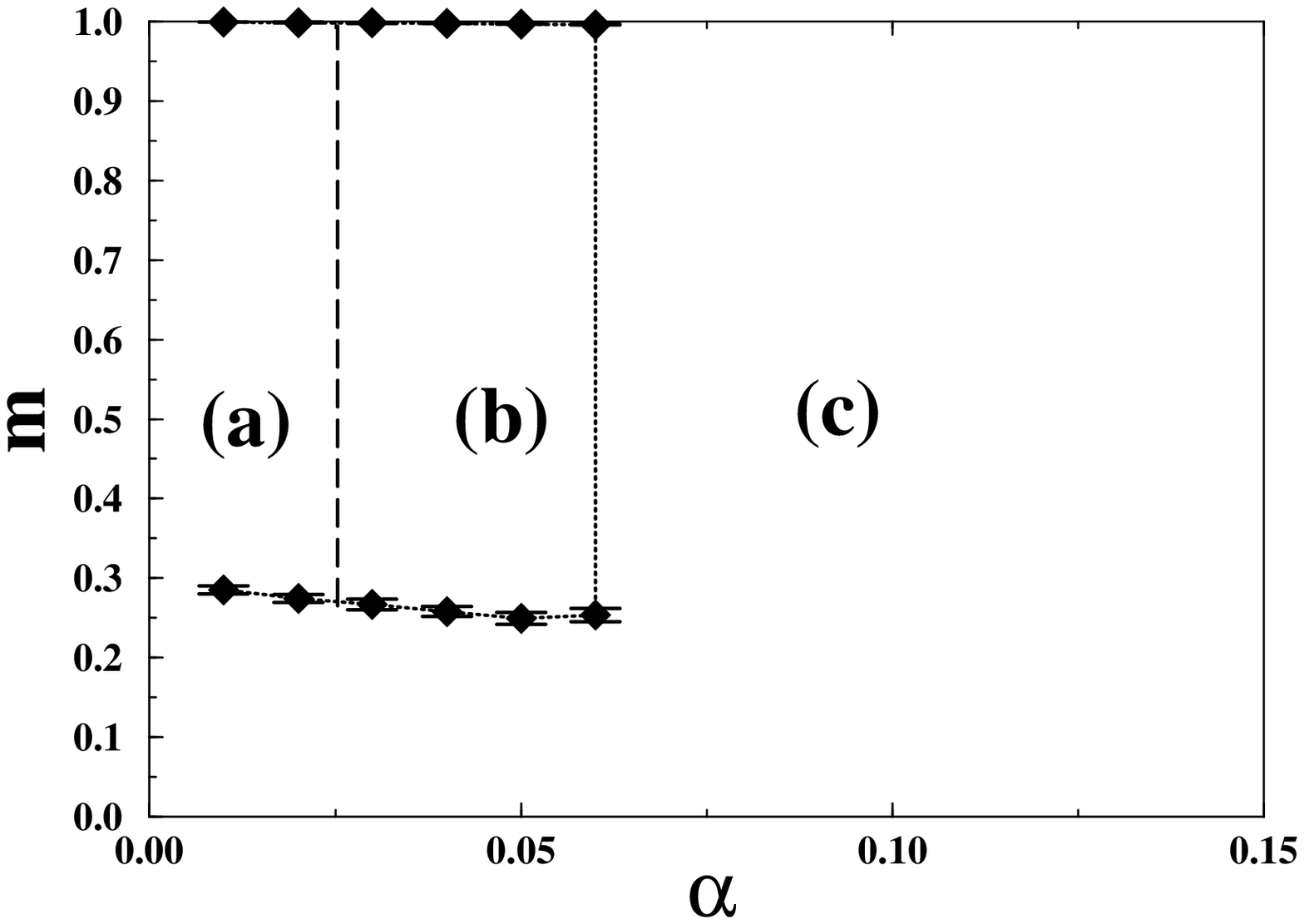}}
    \scalebox{.4}{\includegraphics{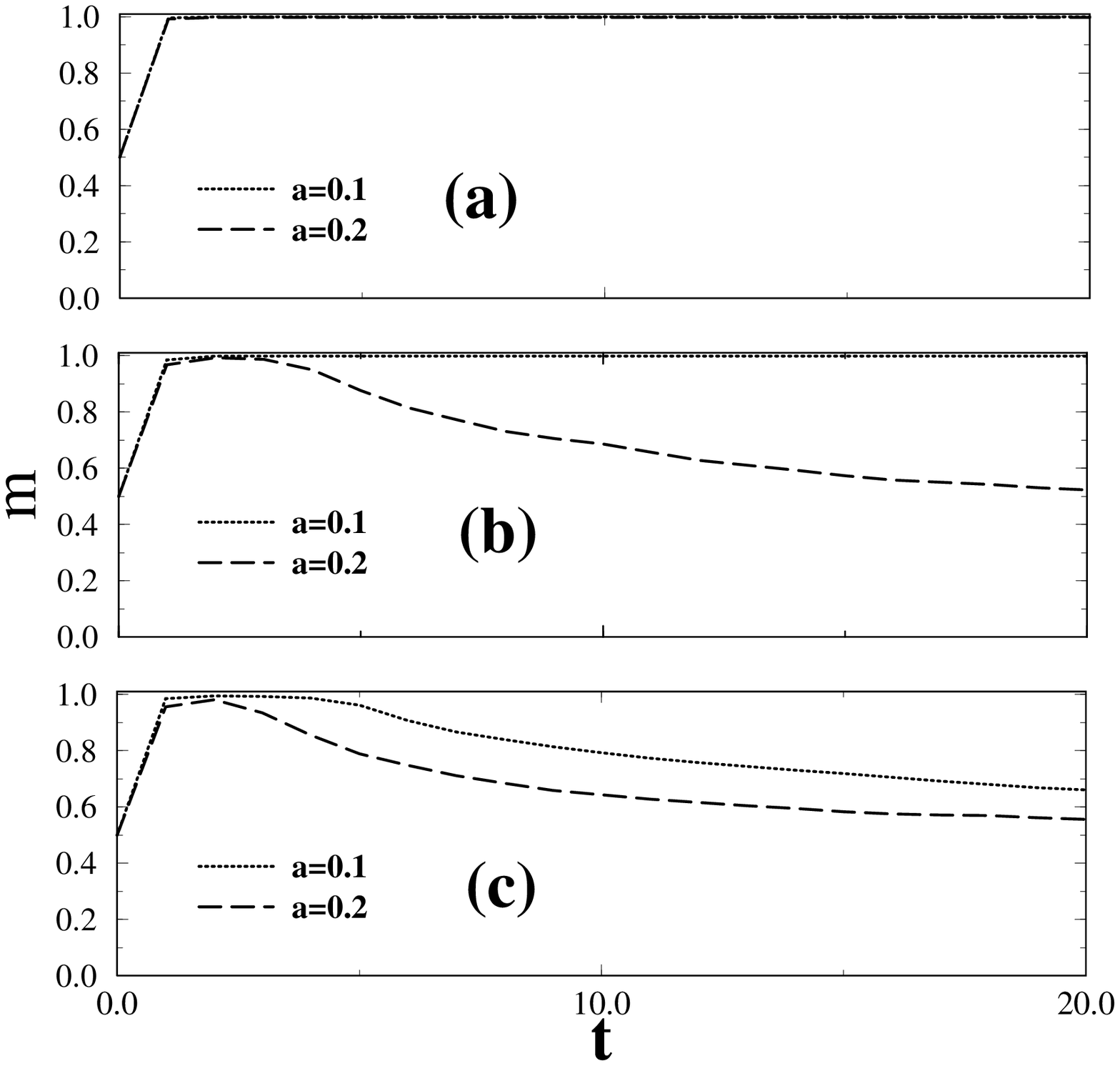}}
    \caption{(left) Basin of attraction for patterns with $a=0.1$ in the case 
      that patterns with $a=0.1$ 
      and $a=0.2$ are embedded in equal number at the same time. Here $H=0.3$.
      The dotted line and the data points represent the theoretical and the 
      numerical results, respectively.
      The vertical long-dashed line indicates the maximum storage capacity for 
      patterns with $a=0.2$. 
      Thus, in the left figure (a) is the region where both activity level 
      patterns are successfully retrieved, 
      (b) is the region where only patterns with $a=0.1$ can be retrieved, 
      and (c) is the region where neither
      can be retrieved.
      (right) Time evolutions of the overlap $m(t)$ for the initial condition 
      $m(0)=0.5$, with $H=0.3$ and 
      $\alpha=0.02,0.05$ and $0.08$, corresponding to the regions (a), (b) and 
      (c) in the left figure, respectively. 
      These are the results of the numerical simulation.
      \label{doublebasin}}
  \end{center}
\end{figure}

Let us consider the case $P_1=P_2$. Note that in this case,
$\alpha_1=\alpha_2$ and the total storage capacity $\alpha$ has the relation
as $\alpha=2\alpha_1=2\alpha_2$. We assume that $a_1=0.1$, $a_2=0.2$ and
$H=0.3$. Under these conditions, the basin of attraction obtained
numerically from the theory is displayed in the left panel of 
Fig.\ref{doublebasin}. We should note that in the region (b), the patterns
with activity $a_1$ can be recalled, while the patterns with activity $a_2$
cannot. Since we consider the case $P_1=P_2$, the vertical lines correspond
to half of the usual maximum storage capacities.
In the right figures, we display typical behavior of the overlap $m(t)$
for the initial condition $m(0)=0.5$.
The load parameters $\alpha=0.02,0.05$ and $0.08$
used here correspond to the regions (a), (b) and (c) shown 
in the left figure, respectively.
In the right panel of Fig.\ref{doublebasin} appear comparisons of the time
evolution of the overlap for two different values of $a$ in three different
regions of $\alpha$. The evolutions corresponding to these values of $a$
represent the retrieval processes of $\xi_i^1$ and $\xi_i^{P_1+1}$ associated
with the activity levels $a_1$ and $a_2$, respectively.
Note that in the region (b), the patterns with $a=0.2$ act only as noise
when the network is recalling the pattern with $a=0.1$.
As a whole, the above findings suggest that the network has a good ability 
to retrieve patterns with different firing rates. We should remark that the
patterns with activity $a_1$ can be stored more stably rather than those with
activity $a_2$ when $a_1< a_2$.

\section{Dilution}
In this section, we study the influence of random synaptic dilution on
the model's associative memory capability. For the case of the phase
oscillator model, that is, when $a=1$ and $H=0$, this effect has
already been reported \cite{aoyagi3,kitano1}. Following the 
method used in that case to treat random synaptic dilution in our model, 
we assume that the synaptic efficacies take the form

\begin{equation}
  \overline{C}_{ij}=\frac{c_{ij}}c C_{ij},
  \label{cutcij}
\end{equation}
where $C_{ij}$ is the standard Hebbian matrix, as defined by Eq.(\ref{cij}),
and, the $c_{ij}$ are independent random variables, taking the values 
$1$ and $0$ with probabilities $c$ and $1-c$, respectively.
Note that the dilution parameter $c$ represents the ratio of 
connected synapses. In the limit $N\rightarrow\infty$, 
Eq.(\ref{cutcij}) can be regarded as
\begin{equation}
  \overline{C}_{ij}=C_{ij}+\eta_{ij},
\end{equation}
where the synaptic noise $\eta_{ij}$ is a complex Gaussian noise with mean
$0$ and variance $\eta^2/N$ \cite{cut}. The relationship between the dilution
parameter $c$ and the variance $\eta$ can be calculated as
\begin{equation}
  \eta^2=\frac{1-c}c\alpha\label{eta}.
\end{equation}
Under the same assumption as that used in obtaining Eq.(\ref{order}), 
we can separate the local
field into the signal and two noise parts as follows:
\begin{eqnarray}
  h_i(t)&=&\sum_j^N\overline{C}_{ij}W_j(t)\nonumber\\
  &=&\sum_j^N(C_{ij}+\eta_{ij})W_j(t)\nonumber\\
  &=&\sum_j^N(\frac1{aN}\sum_{\mu=1}^P\xi_i^\mu\tilde{\xi_j^\mu}
    +\eta_{ij})W_j(t)\nonumber\\
  &=&M^1(t)\xi_i^1+z_i(t)\\
  &=&M^1(t)\xi_i^1+z_i^c(t)+z_i^s(t),
\end{eqnarray}
where $z_i^c(t)$ and $z_i^s(t)$ are defined as
\begin{eqnarray}
  z_i^c(t)&=&\frac1{aN}\sum_j^N\sum_{\mu\neq1}^P\xi_i^\mu
  \tilde{\xi_j^\mu}W_j(t)\\
  z_i^s(t)&=&\sum_j^N\eta_{ij}W_j(t).
\end{eqnarray}
The new noise term $z_i^s(t)$ is caused by synaptic dilution, while the
term $z_i^c(t)$ is like the noise defined by Eq.(\ref{z_i}), representing 
the crosstalk noise arising from the other embedded patterns. 
In analogy to our treatment in Sec.\ref{onn}, we assume that $z_i^c(t)$ and
$z_i^s(t)$ are independent complex Gaussian noises with mean $0$ and variance
$\sigma_c^2(t)$ and $\sigma_s^2(t)$, respectively. Therefore, $z_i(t)$ 
can be regarded as a complex Gaussian noise with mean $0$ and variance
$2\sigma^2(t)=\sigma_c^2(t)+\sigma_s^2(t)$. Thus, we can 
derive dynamical equations in the same way as in Sec.\ref{onn}.
The form of the macroscopic order parameter $m(t+1)$ is the same as 
that given by Eq.(\ref{m}):
\begin{eqnarray}
  m(t+1)&=&\frac1{aN}\sum_j|\xi_j^1|f(||\xi_j^1|m(t)+z_j(t)|)
  \frac{|\xi_j^1|m(t)+z_j(t)}{||\xi_j^1|m(t)+z_j(t)|}
  \nonumber\\
  &=&\left<\left<f(|m(t)+z(t)|)\frac{m(t)+z(t)}{|m(t)+z(t)|}
    \right>\right>_{z(t)}.
\end{eqnarray}

\begin{figure}
  \begin{center}
    \scalebox{.4}{\includegraphics{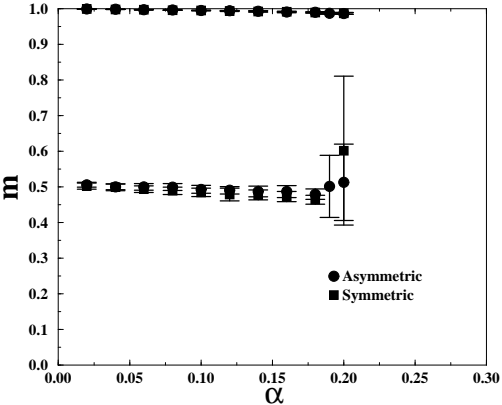}}
    \caption{Comparison of basins for symmetrical and asymmetrical dilution.
      Both were obtained from numerical simulations with $N=2000$ for $20$ 
      trials.\label{sym-asym}}
  \end{center}
\end{figure}

First, applying the theory of statistical neurodynamics to $z_i^s(t)$, we
calculate $\sigma_s^2(t)$. However, to apply the theory of statistical
neurodynamics, it is necessary to take into account the second term in
\begin{equation}
  z_i^s(t+1)=\sum_j^N\eta_{ij}f(|h_j^0|)\frac{h_j^0}{|h_j^0|}+
  W_i(t)\sum_j^N\eta_{ij}\eta_{ji}
  \left(\frac{f^{\prime}(|h_j^0|)}2+\frac{f(|h_j^0|)}{2|h_j^0|}\right),
\end{equation}
where $h_j^0=h_j(t)-\eta_{ji}W_i(t)$.
As reported in \cite{kitano1}, for the asymmetrical case,
$\eta_{ij}\neq\eta_{ji}$, the second term
here can be ignored. Then, as seen from Fig.(\ref{sym-asym}), there is little
difference between symmetric and asymmetric cases in a sparse coding system.
Thus we conjecture that this term can be ignored altogether. Doing so, 
in the $N\rightarrow\infty$ limit, we obtain
\begin{eqnarray}
  \sigma_s^2(t)&=&\eta^2Q(m(t),\sigma(t))\nonumber\\
  &=&\frac{1-c}c\alpha Q(m(t),\sigma(t)),
\end{eqnarray}
where $Q(m(t),\sigma(t))$ is defined by Eq.(\ref{2ndq}).

For the crosstalk noise $z_i^c(t)$, in analogy to Eq.(\ref{z2}), we obtain
\begin{equation}
  z_i^c(t+1)=K(m(t),\sigma(t))+z_i^c(t)G(m(t),\sigma(t)),
  \label{zc}
\end{equation}
where $K(m(t),\sigma(t))$ and $G(m(t),\sigma(t))$ are defined by
Eqs.(\ref{k}) and (\ref{g}), respectively.

\subsection{Equilibrium State}
\begin{figure}
  \begin{center}
    \scalebox{.4}{\includegraphics{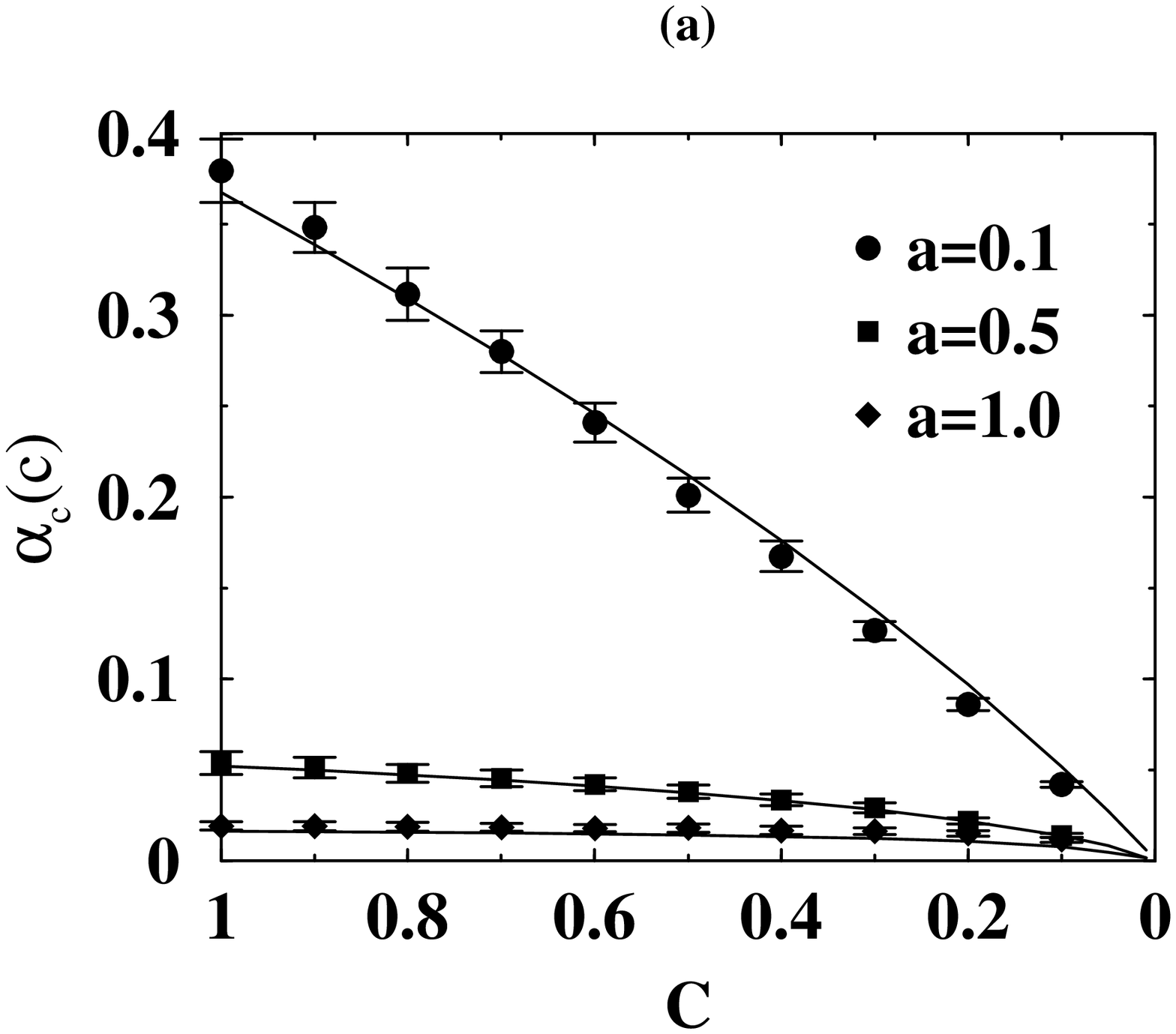}\includegraphics{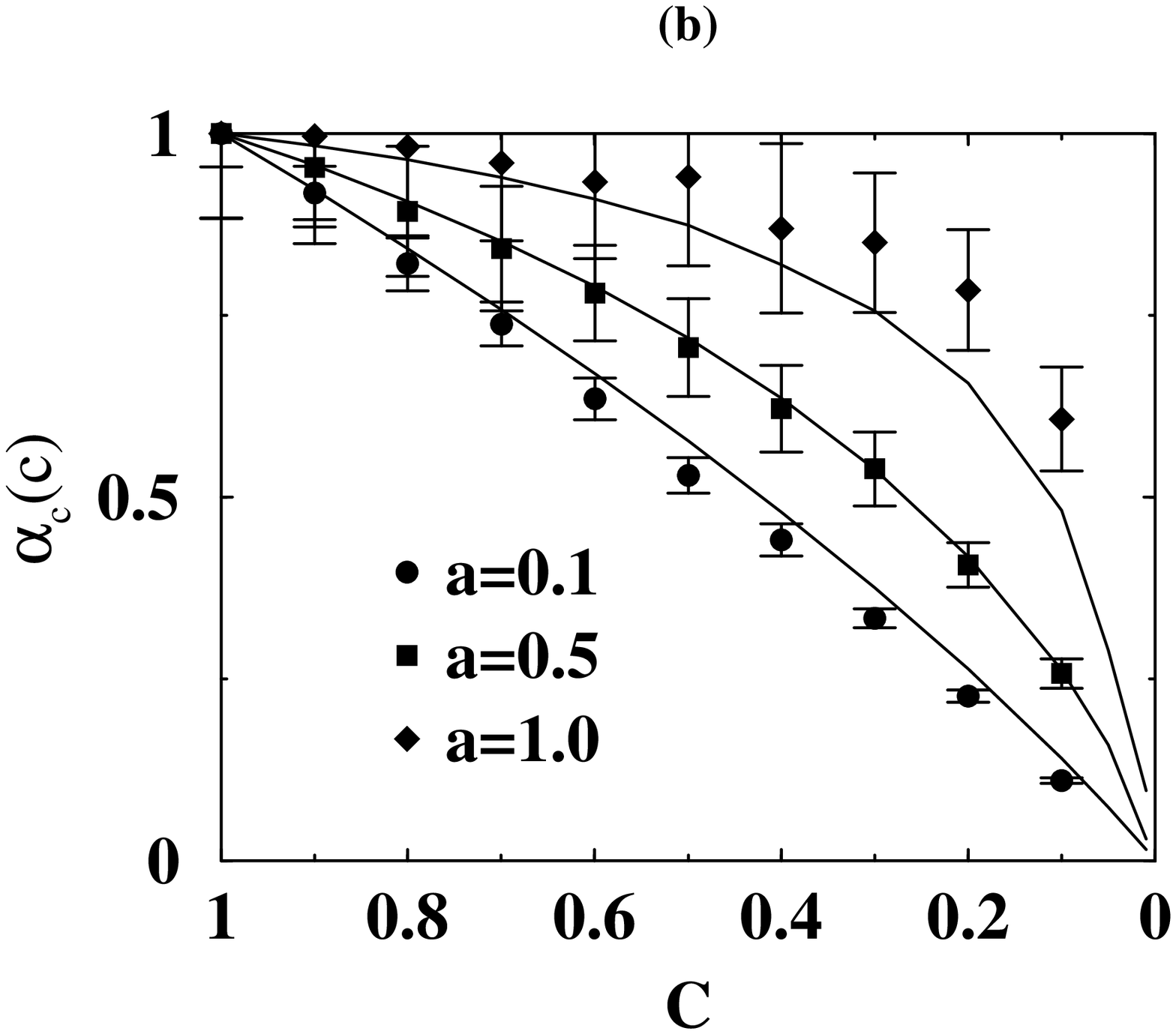}}
    \scalebox{1.0}{\includegraphics{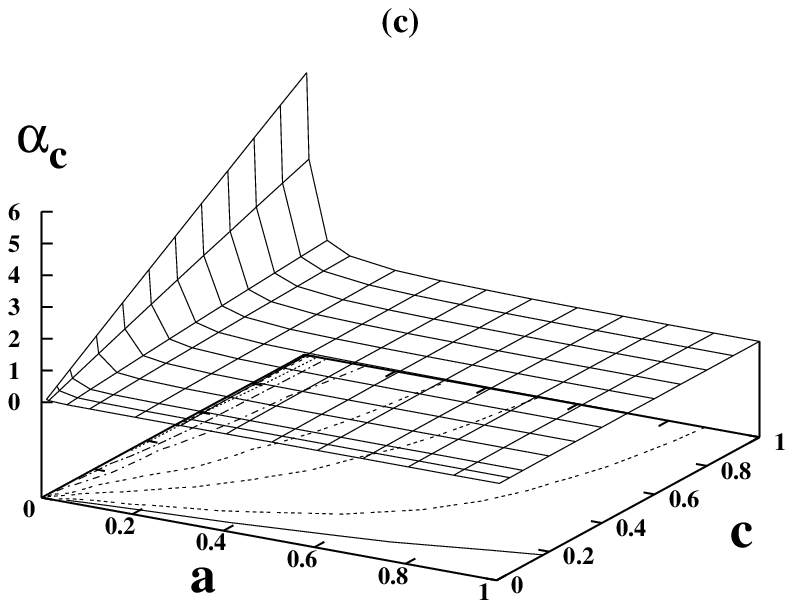}}
    \caption{(a) Dependence of the maximum storage storage capacity $\alpha_c$ 
      on the ratio of
      connectivity $c$ for various firing rates with $H=0.5$. The solid lines 
      represent the 
      theoretical results, and the data points represent the numerical results 
      with $N=2000$ 
      for $20$ trials. (b) Same as (a), but normalized by 
      $\alpha_c(c)/\alpha_c(1)$.
      (c) A three-dimensional isometric plot of the maximum storage capacity 
      $\alpha_c$
      as a function of $a$ and $c$ for $H=0.5$. This was obtained from the 
      theoretical results.
      \label{dilution1}}
  \end{center}
\end{figure}

In the equilibrium state, putting $z_i^c(t)=z_i^c(t+1)$ into Eq.(\ref{zc}),
we obtain
\begin{eqnarray}
  z_i^c&=&\frac K{1-G}\\
  \sigma_c^2&=&\frac{\alpha Q}{(1-G)^2}.
\end{eqnarray}
Therefore, $\sigma^2$, the variance of $z(t)$, takes the form
\begin{eqnarray}
  \sigma^2&=&\frac12\sigma_c^2+\frac12\sigma_s^2\nonumber\\
  &=&\left(\frac1{2(1-G)^2}+\frac{1-c}{2c}\right)\alpha Q.
  \label{sigmaconst}
\end{eqnarray}
Consequently, the properties of the network in the equilibrium state can be
calculated from Eqs.(\ref{mconst}), (\ref{sigmaconst}), (\ref{gconst}) and
(\ref{qconst}). In Fig.\ref{dilution1}, we summarize the theoretical results
concerning the dependence on the ratio of connected synapses $c$. In
Figs.\ref{dilution1}(a) and (b), it is found that the maximum storage
capacity is an increasing function of the connectivity $c$. Particularly,
we can see from (b) that, as the activity level $a$ becomes small,
the storage capacity decreases almost linearly with the connectivity
$c$. A similar linear dependence is observed in the case of the diluted
Hopfield model. On the other hand, in (c), it is found that even if the
ratio of connectivity is small, the maximum storage capacity tends to
diverge in the limit $a\rightarrow0$. However, it seems that the rate
of this divergence decreases as $c$ decreases.

\subsection{Non-equilibrium State}
\begin{figure}
  \begin{center}
    \scalebox{.4}{\includegraphics{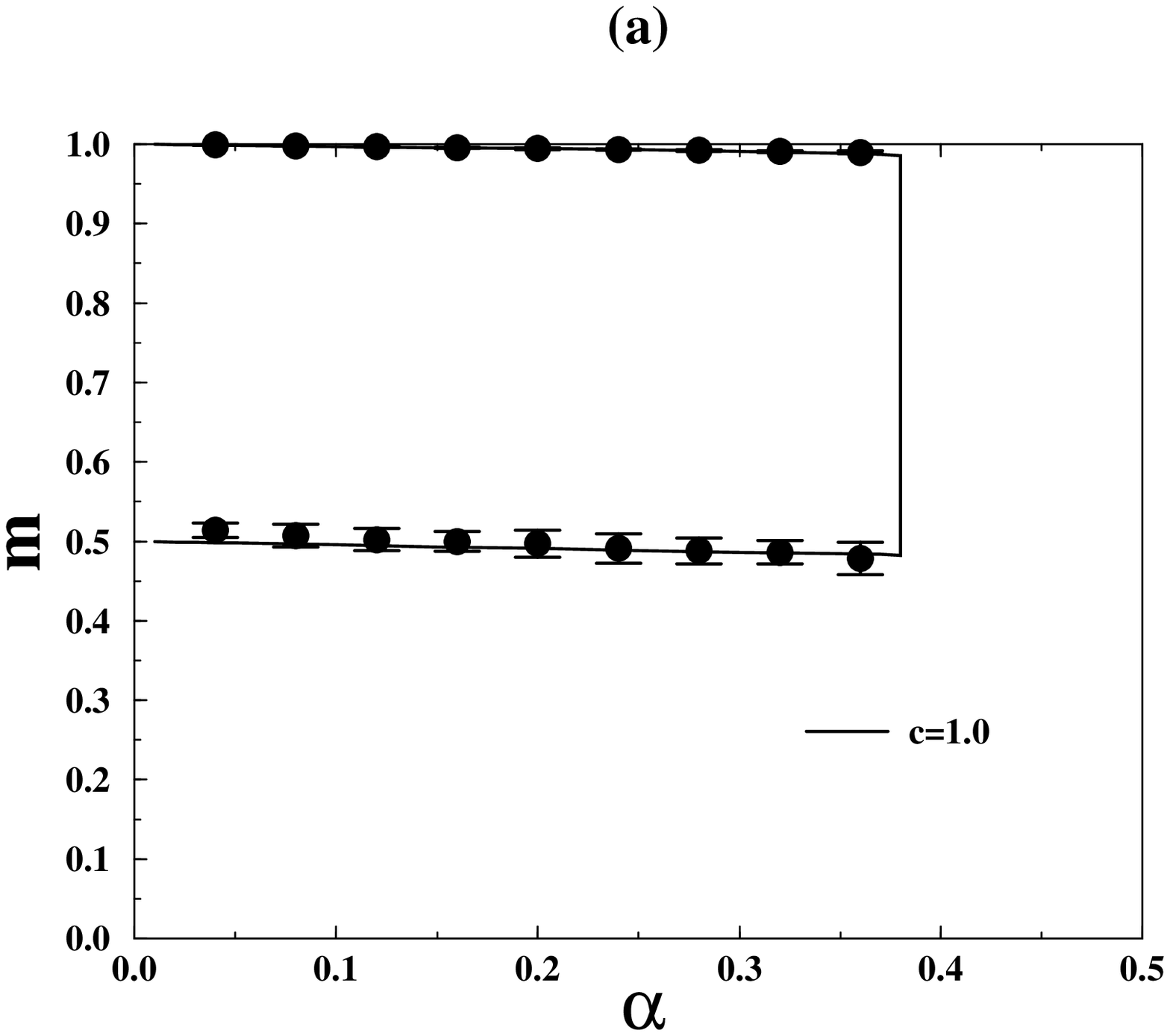}
      \includegraphics{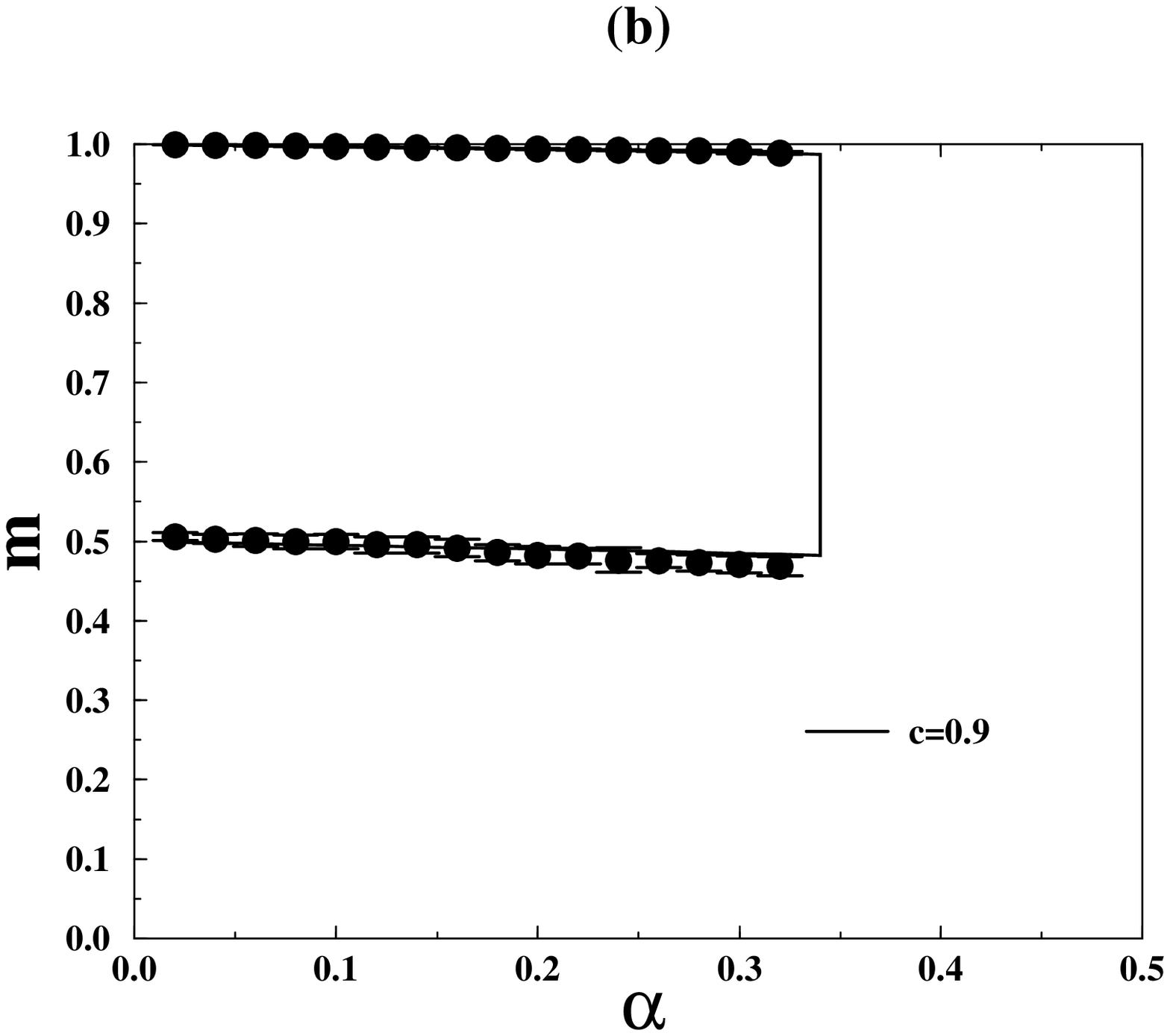}}
    \scalebox{.4}{\includegraphics{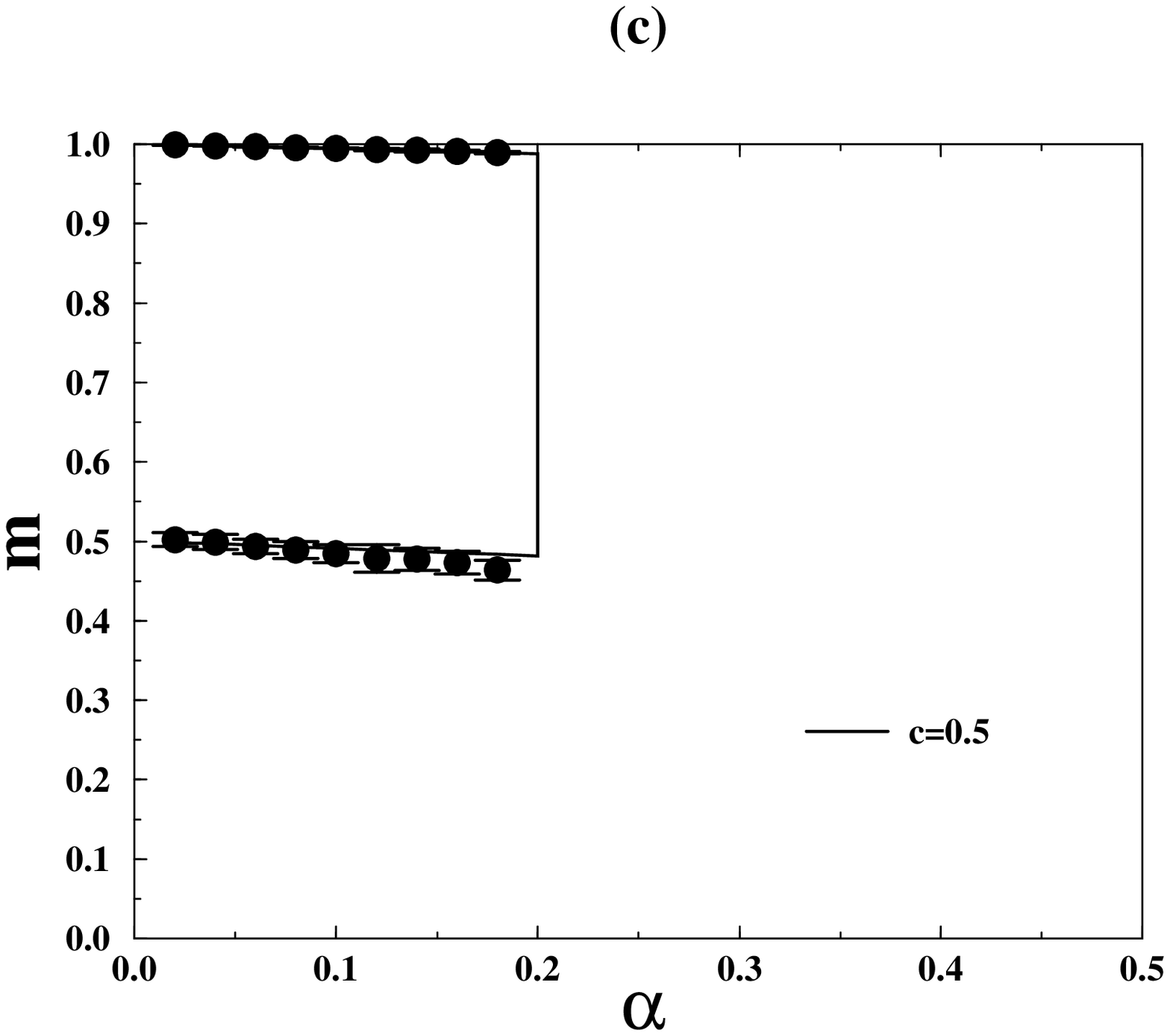}
      \includegraphics{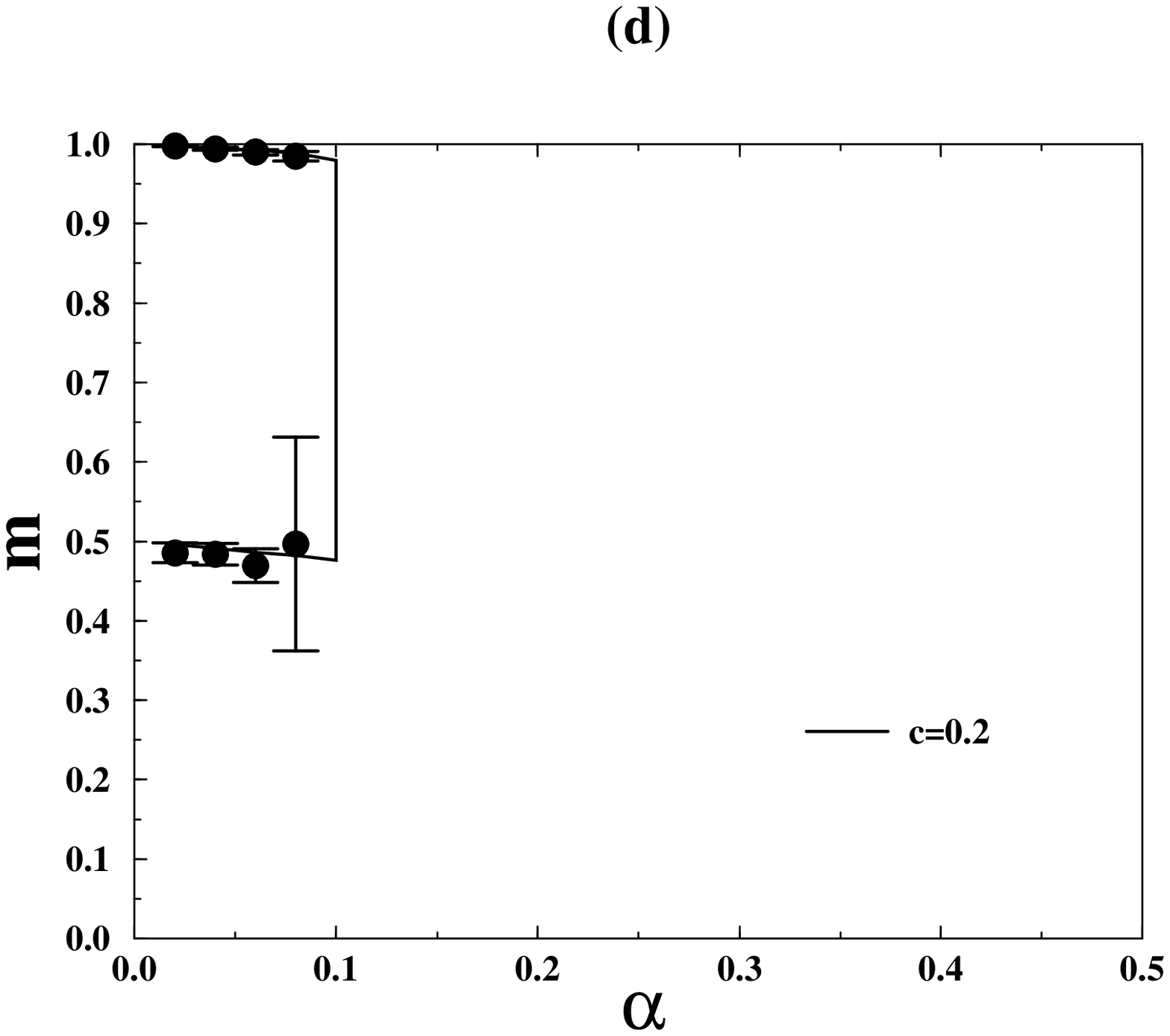}}
    \caption{Basins of attraction for various values of the connectivity $c$. 
      The solid lines represent the theoretical results, and the data
      points represent the numerical results with $N=2000$ for $20$ trials.
      \label{dilution2}}
  \end{center}
\end{figure}

In order to estimate the robustness with respect to synaptic damage, we
should also consider the influence of synaptic dilution on the retrieval
process, particularly, on the basin of attraction. For this purpose,
we can apply the same method as used in the derivation of
Eq.(\ref{2ndsigma}). After some calculations, the resulting recursion
equations at second order are given by
\begin{eqnarray}
  \sigma^2(t+1)&=&\frac{\alpha}2 Q(m(t),\sigma(t))
  +\sigma^2(t)G^2(m(t),\sigma(t))\nonumber\\
  &&+\alpha G(m(t),\sigma(t))X(t+1,t)\nonumber\nonumber\\
  &&+\alpha a^2m(t+1)m(t-1)G(m(t),\sigma(t))G(m(t-1),\sigma(t-1))\nonumber\\
  &&+\frac12(1-G^2(m(t),\sigma(t)))\eta^2Q(m(t),\sigma(t))\\
  X(t+1,t)&=&Re<W_j(t+1)\tilde{W}_j(t)>_j\label{xt2}\\
  \rho(t,t-1)&=&\frac{\alpha X(t,t-1)}{2\sigma(t)\sigma(t-1)}
  +\frac{\sigma(t-1)}{\sigma(t)}G(m(t-1),\sigma(t-1))\label{ro2}\\
  &&+\frac{X(t,t-1)-G(m(t-1),\sigma(t-1))Q(m(t-1),\sigma(t-1))}
  {2\sigma(t)\sigma(t-1)}\eta^2,\nonumber
\end{eqnarray}
where $\eta$ is related to the connectivity $c$ via Eq.(\ref{eta}).
As mentioned in Sec.\ref{onn}, Eq.(\ref{ro2}) is used to calculate
Eq.(\ref{xt2}). For initial conditions, we adopt the values used
in Sec.\ref{onn}, except in the present case we use $\sigma^2(0)=a\alpha/2c$.
In the case of $a=0.1$ and $H=0.5$, Fig.\ref{dilution2} illustrates
the theoretical results concerning the basins of attraction for various values
of the connectivity $c$. We can see that near saturation $\alpha\sim\alpha_c$,
the basin of attraction remains large even for low connectivity.
Therefore, we find that synaptic dilution has little influence on the width of
the basin of attraction, even though the storage capacity decreases with
the connectivity.

\section{Conclusion}
In this paper, we have presented a simple extended model of oscillator
neural networks to allow for the description of the non-firing state.
We have studied the model's associative memory capability for
sparsely coded phase patterns, in which some neurons are in the non-firing
state and the other neurons encode information in the phase variable 
representing the timing of neuronal spikes. 
In particular, applying the theory of statistical neurodynamics, we 
have evaluated the maximum storage capacity and derived the basin of attraction.
We have found the following properties of our model in its basic form:
\begin{itemize}
\item The storage capacity diverges as the activity level decreases to zero.
  It was numerically found that the storage capacity diverges proportionally 
  with $1/a\ln a$ in the limit $a\rightarrow0$.
\item Even just below the maximum storage capacity, the basin of
  attraction remains large.
\end{itemize}
We then investigated the model with regard to the size of the basin of
attraction. We found that with the model in its basic form, the basin of
attraction is smaller than those of the binary model and the phase oscillator.
For associative memory, it is desirable that the basin of
attraction be large. For this reason, we considered employing
a dynamically adjusted threshold, and we found the following:
\begin{itemize}
\item The basin of attraction can be enlarged by using the dynamically
  adjusted threshold.
\end{itemize}
In view of biology, the neurons may die of age or be injured by accident.
Thus, the robustness with respect to synaptic damage is important for real
neuronal systems. For this reason, we also investigated our model with regard
to robustness, and we found the following:
\begin{itemize}
\item It was found that the system is robust with respect to synaptic damage:
  Even in the case of a high cutting rate, the basin of attraction remains
  large, and the maximum of the storage capacity diverges in the $a\rightarrow0$
  limit. For low activity patterns, the maximum storage capacity decreases
  almost linearly with the ratio of connected synapse.
\end{itemize}
The above properties are common with the Hopfield model. In addition, we have
found that our model possesses a novel feature not seen in the Hopfield model.
In realistic situations, the activity level of the firing pattern may
generally depend on the content of the information. Using traditional
neural network models based only on firing rate coding, however,
we encounter difficulties when the network simultaneously stores such
patterns with different activity levels. Contrastingly, with our model,
we found the following:
\begin{itemize}
\item Unlike the Hopfield model, provided that the phase distribution
  in the embedded patterns is uniform, it was shown that patterns with
  different activity levels can be memorized simultaneously.
\end{itemize}

In conclusion, from the above findings it is seen that the oscillator
neural network exhibits good performance in the sparse coding situation.
Therefore, we believe that these results support the plausibility of
temporal coding and we hope that they encourage attempts for more detailed
explorations of the nature of temporal coding. As a future work, we wish
to consider one of the applications of the oscillator model. In this paper
we considered coherent oscillations of only a uniform distribution. It is more
interesting, however, to consider systems that are not so strongly constrained.
For example, neurons could be organized into ensembles characterized
by different phase values. In such a situation, each ensemble would carry out
a different function independently, and the neural system as a whole could
carry out several tasks simultaneously. Using this kind of synchrony,
we may obtain a simple solution to the binding problem.

\section{Acknowledgement}
We express our gratitude to Prof. T. Munakata and Dr. K. Kitano for helpful
discussions. This work was supported by the Japanese Grant-in-Aid for
Science Research Fund from the Ministry of Education, Science and Culture.

\newpage
\newpage
\newpage
\newpage
\newpage
\end{document}